\documentclass[a4paper,aps,prd,showpacs,showkeys,twocolumn,10pt]{revtex4-1}
\usepackage{graphicx,slashed,amssymb,subfigure}
\usepackage{amsmath,bm}
\def\be{\begin{eqnarray}}
\def\ee{\end{eqnarray}}
\def\nn{\nonumber\\}

\def\({\left(}
\def\){\right)}
\newcommand*{\mycdot}{\kern-.15em\cdot\kern-.15em}

\usepackage[colorlinks=true,linktocpage=true,linkcolor=blue,citecolor=blue]{hyperref}
 \allowdisplaybreaks

\begin{abstract}
	
\end{abstract}
\begin{document}

\title{Quark mass dependent collective excitations and quark number susceptibilities within the hard thermal loop approximation}

\author{Najmul Haque}
\email{Najmul.Haque@theo.physik.uni-giessen.de}
\affiliation{Institut f\"ur Theoretische Physik, Justus-Liebig--Universit\"at Giessen, 35392 Giessen, Germany}

\begin{abstract}
 We calculate all those QCD $N$-point functions which are relevant for a three-loop QCD thermodynamics calculation with finite quark masses within the hard thermal loop approximation. Using the effective quark propagator, we also calculate second-order quark and baryon number susceptibilities within the hard thermal loop approximation and compare the results with available lattice data.
\end{abstract}	
\maketitle

\section{Introduction}
It is an experimental fact that colored quarks and gluons are confined to hadrons by the strong interaction. The theory, which correctly describes this interaction, is known as quantum chromodynamics (QCD). In extreme conditions, such as at very high temperatures and/or densities, hadronic states are believed to undergo a (partially crossover) transition into a deconfined state of quarks and gluons,  known as quark-gluon plasma (QGP). Such extreme conditions existed in the very early Universe and can also be generated in ultrarelativistic heavy-ion collision experiments at the Relativistic Heavy Ion Collider (RHIC), the Large Hadron Collider (LHC), and in the future also the Facility for Antiproton and Ion Research. In addition, the cores of the densest astrophysical objects in existence, neutron stars, may contain cold deconfined matter, commonly referred to as quark matter. The reason why deconfined matter is expected to be encountered at high energy densities is related to the asymptotic freedom of QCD, i.e. the fact that the value of the strong coupling constant decreases logarithmically as a function of the energy scale. Proceeding to higher energies, nonperturbative effects are also expected to diminish in importance, and calculations based on a weak coupling expansion should eventually become feasible at such extreme conditions. This is very important for a successful quantitative description of the system especially at nonzero chemical potentials, as no nonperturbative first principles method applicable to finite-density QCD exists due to the sign problem of lattice QCD. Unfortunately, at energy densities of phenomenological relevance for most practical applications, the value of the strong coupling constant is not small, and it in fact turns out that, e.g. for bulk thermodynamic quantities a strict expansion in the QCD coupling constant converges only at astronomically high temperatures and chemical potentials. The source of this problem has been readily identified as the infrared sector of the theory, i.e. contributions from soft gluonic momenta of the order of the Debye mass or smaller. This suggests that, to improve the situation, one needs a way of reorganizing the perturbative series that allows for the soft contributions to be included in the result in a physically consistent way. To date, two successful variations of the resummed perturbation theory have been introduced as a remedy, namely, the hard thermal loop perturbation theory (HTLpt)~\cite{Braaten:1989mz,Andersen:1999fw,Haque:2013sja,Haque:2014rua,Haque:2012my,Haque:2013qta,Andersen:2015eoa} and the dimensionally reduced effective field theory~\cite{Kajantie:2002wa,Vuorinen:2003fs}, of which the latter is applicable only to high temperatures but the former, in principle, covers also the cold and dense part of the QCD phase diagram. Both approaches have been shown to lead to results which are in quantitative agreement with lattice QCD in the high-temperature and zero (or small-) density limit, down to temperatures a few times the pseudocritical temperature of the deconfinement transition.

In nearly all thermodynamic calculations applying the perturbation theory, either the temperature or the chemical potentials are assumed to be the dominant energy scale in the system and, in particular, much larger than the QCD scale or any quark masses. It is, however, questionable whether the latter is necessarily a good approximation at the lowest temperatures and densities where the HTLpt results are typically applied; the strange quark mass is after all of the order of 100 MeV, which is certainly not negligible at the temperatures reached in the RHIC and LHC heavy-ion experiments. The effects of quark masses on the equation of state may not be very large, but they will be much more visible in quark
number susceptibilities. For example, in a massless approximation, one cannot distinguish (i) second-order quark number susceptibility (QNS) from three times the second-order baryon number susceptibility (BNS) and (ii) heavy-light fourth-order off-diagonal susceptibility $(\chi_{uuss})$ from the light-light one $(\chi_{uudd})$. In the presence of finite quark masses, one may able to distinguish those susceptibilities as being done in a lattice QCD framework.

There have been very few high-order perturbative calculations incorporating finite quark masses, and, in the context of high temperatures, the state-of-the-art result is from a two-loop perturbative calculation performed long ago~\cite{Kapusta:1979fh} in a renormalization scheme for quark masses. Later, the two-loop perturbative calculation for the thermodynamical quantities has been extended in Refs.~\cite{Laine:2006cp,Graf:2015tda} using the $\overline{\rm MS}$ renormalization scheme. Additionally, the current state-of-the-art result for the QCD pressure at a zero temperature and finite quark mass is from a three-loop bare perturbation theory~\cite{Kurkela:2009gj}. On the other hand, finite quark masses have not been considered at all in any resummed perturbative framework for thermodynamic calculations.

Collective excitations of heavy fermion have been studied long ago in Ref.~\cite{Petitgirard:1991mf} calculating the heavy-fermion propagator. Later, in Ref.~\cite{Seipt:2008wx}, the authors have studied quark mass-dependent thermal excitations calculating both quark and gluon propagators. In this article we calculate all the QCD $N$-point functions, such as the gluon propagator, quark propagator, and three- and four-point quark-gluon vertices considering the HTL approximation. These $N$-point functions are relevant for higher order thermodynamics calculation within  the HTLpt framework. In this direction, we calculate QNS and BNS using the two-loop $\Phi$-derivable self-consistent approximation within the HTL scheme, as described in Refs.~\cite{Blaizot:2000fc,Blaizot:2001vr} for massless quarks. 

The paper is organized as follows. In Sec.~\ref{sec:quark_prop}, we calculate the HTL effective quark propagator with a finite quark mass. In Sec.~\ref{sec:3vertex}, we discuss the three- and four-point quark-gluon vertices with a finite quark mass. In the next section (Sec.~\ref{sec:qns}), we calculate the second-order light quark as well as strange quark number susceptibility which lead to the computation of the second-order BNS. In Sec.~\ref{sec:conclusion}, we conclude our results. In the appendix we also discuss the HTL effective gluon propagator and gluon dispersion relations including finite quark masses.

\section{Quark propagator}
\label{sec:quark_prop}
The free massive quark propagator is
\be
S_0(P) &=& \frac{i}{\slashed{P}-m_f}. \label{free_prop} 
\ee
The inverse of the propagator in Eq.~(\ref{free_prop}) can be written as
\be
i S_0^{-1}(P) &=& \slashed{P}-m_f. \label{free_prop_inv}
\ee
Now, the inverse of the in-medium quark propagator is
\be
i S^{-1}(P) &=& i  S_0^{-1}(P)- \Sigma(P), \label{med_prop_inv_def}
\ee
where $\Sigma(P)$ is the one-loop quark self-energy and can be decomposed as
\be
\Sigma(P)=-a\slashed{P}-b\slashed{u} + cm_f.\label{sigma_decomp}
\ee
The coefficients $a,b,$ and $c$ can be obtained, respectively, as
\be
a&=&-\frac{1}{4p}\mbox{Tr}[\bm{\gamma\cdot\hat{p}}\ \Sigma(P)],\nn
b&=&-\frac{1}{4}\mbox{Tr}[\gamma_0\ \Sigma(P)]+\frac{p_0}{4p}\mbox{Tr}[\bm{\gamma\cdot\hat{p}}\ \Sigma(P)],\nn
c&=&\frac{1}{4m_f}\mbox{Tr}[\Sigma(P)].
\label{abc}
\ee
So, Eq.~(\ref{med_prop_inv_def}) becomes
\be
i S^{-1}(P) &=& (1+a)\slashed{P}+b\slashed{u} - (1+c)m_f,\nn
-i S(P) &=& \frac{(1+a)\slashed{P}+b\slashed{u} + (1+c)m_f}{D(p_0,p)},\label{med_prop_inv_decomp}
\ee
where
\be
&&\hspace{-1cm}D(p_0,p)\nn
&=&(1+a)^2P^2 + b^2 - (1+c)^2m_f^2 +2 b(1+a)p_0\nn
% &=& \left[(1+a)p_0 + b -\sqrt{(1+a)^2p^2 + (1+c)^2 m_f^2}\right]\nn
% &\times&\left[(1+a)p_0 + b +\sqrt{(1+a)^2p^2 + (1+c)^2 m_f^2}\right]\nn
&=& D_+(p_0,p)D_-(p_0,p),
\ee
with
\be
&&\hspace{-1cm}D_\pm(p_0,p)\nn &=& (1+a)p_0 + b \mp\sqrt{(1+a)^2p^2 + (1+c)^2 m_f^2}\ .
\ee
Now, the numerator of Eq.~(\ref{med_prop_inv_decomp}) can be written as
\be
&&(1+a)\slashed{P}+b\slashed{u}+(1+c)m_f \nn
&=&\frac{D_+(p_0,p)}{2}\left[\gamma_0+\frac{\bm{\gamma\cdot\hat{p}}(1+a)p+(1+c)m_f}{\sqrt{(1+a)^2p^2 + (1+c)^2 m_f^2}}\right]\nn
&+&\frac{D_-(p_0,p)}{2}\left[\gamma_0-\frac{\bm{\gamma\cdot\hat{p}}(1+a)p+(1+c)m_f}{\sqrt{(1+a)^2p^2 + (1+c)^2 m_f^2}}\right].\quad
\ee 
So, Eq.~(\ref{med_prop_inv_decomp}) becomes
\be
&&\hspace{-0.8cm} -i S(P) \nn&=&\frac{1}{2D_+(p_0,p)}\!\left[\!\gamma_0-\frac{\bm{\gamma\cdot\hat{p}}(1+a)p+(1+c)m_f}{\sqrt{(1+a)^2p^2 + (1+c)^2 m_f^2}}\right]\nn
&\!+\!\!\!&\frac{1}{2D_-(p_0,p)}\!\left[\!\gamma_0+\frac{\bm{\gamma\cdot\hat{p}}(1+a)p+(1+c)m_f}{\sqrt{(1+a)^2p^2 + (1+c)^2 m_f^2}}\right]\!.\quad
\label{med_prop_decomp}
\ee
Zeros of the denominator $[D_\pm(p_0,p)]$ of Eq.~(\ref{med_prop_decomp}) give the dispersion relation.
% % % % % % % % % % % 
\noindent Now,
\be
&&\hspace{-0.5cm}i S^{-1}(P) \nn
&=& (1+a)\slashed{P}+b\slashed{u} - (1+c)m_f\nn
&=&[(1+a)p_0+b]\gamma_0 - (1+a)\bm{\gamma\cdot\hat{p}}\ p-(1+c)m_f\nn
&=& A_0(p_0,p)\gamma_0 - A_s \bm{\gamma\cdot\hat{p}} - A_m,
\ee
with
\be
A_0(p_0,p)&=&(1+a)p_0+b,\nn
A_s(p_0,p)&=&(1+a)p,\nn
A_m(p_0,p)&=&(1+c)m_f.
\label{A0sm_def}
\ee
%%%%%%%%%%%%%%%%%%%%%%
Alternatively,
\be
D_\pm(p_0,p) &=& A_0(p_0,p) \nn
&&\hspace{0cm}\mp \sqrt{A_0(p_0,p)^2 + A_m(p_0,p)^2}.
\label{Dpm_def}
\ee
The self-energy can also be written using Eqs.~(\ref{sigma_decomp}) and~(\ref{A0sm_def}) as
\be
\Sigma(P)&=&-a\slashed{P}-b\slashed{u} + cm_f\nn
&=&-(ap_0+b)\gamma_0+ a p \bm{\gamma\cdot\hat{p}} + cm_f\nn
&=&[p_0-A_0(p_0,p)]\gamma_0 +[A_s(p_0,p)-p]\bm{\gamma\cdot\hat{p}}\nn
&&+[A_m(p_0,p)-m_f].
\ee
Equation~(\ref{med_prop_decomp}) can also be written using Eq.~(\ref{A0sm_def}) as
\be
&&-i S(P)\nn
 &=&\frac{1}{2D_+(p_0,p)}\left[\gamma_0-\frac{A_s(p_0,p)\bm{\gamma\cdot\hat p}+A_m(p_0,p)}{\sqrt{A_s(p_0,p)^2 + A_m(p_0,p)^2}}\right]\nn
&+&\frac{1}{2D_-(p_0,p)}\left[\gamma_0+\frac{A_s(p_0,p)\bm{\gamma\cdot\hat p}+A_m(p_0,p)}{\sqrt{A_s(p_0,p)^2 + A_m(p_0,p)^2}}\right]\!.\quad
\label{med_prop_decomp1}
\ee
\begin{figure}[tbh]
	\includegraphics[scale=0.45]{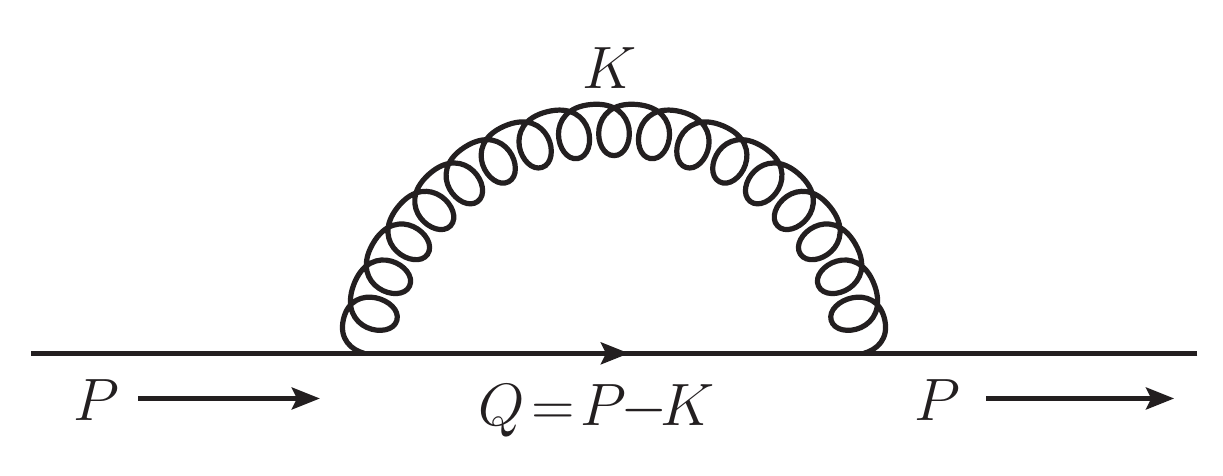}
	\caption{Quark self-energy}
	\label{fig:quark_self}
\end{figure}
%%%%%
\noindent Now, the quark self-energy in the Feynman gauge can be obtained from the Feynman diagram in Fig.~\ref{fig:quark_self} as 
\be
-i\Sigma(P) &=& (ig)^2C_F \int\frac{d^4K}{(2\pi)^4}~\gamma_\mu \frac{i}{\slashed{P}-\slashed{K}-m_f} \gamma_\nu \frac{-ig^{\mu\nu}}{K^2}\nn
\Sigma(P)&=& 2g^2C_F T\sum_{n}\int\frac{d^3k}{(2\pi)^3}~ [\slashed{K}+2m_f]\nn
&\times& \Delta(K)\tilde{\Delta}(P-K),
\label{quark_self}
\ee
% % % % % 
where $\Delta(K)$ is the gluon propagator whereas $\tilde{\Delta}(P-K)$ represents the fermionic propagator and both can be written as
\be
\Delta(K)&=&\frac{1}{K^2}=\frac{1}{k_0^2-k^2},\nn
\tilde{\Delta}(P-K)&=&\frac{1}{(P-K)^2-m_f^2}\nn
&=&\frac{1}{(p_0-k_0)^2-|\bm{p-k}|^2-m_f^2}.
\ee
To compute the quark self-energy in Eq.~(\ref{quark_self}), we need to compute the following Matsubara sums: 
$T\sum_{k_0}\tilde{\Delta}(P-K)\Delta(K)$ and
$T\sum_{k_0}k_0\tilde{\Delta}(P-K)\Delta(K)$. 
% % % % % 
We calculate them using the mixed representation prescribed by Pisarski~\cite{Pisarski:1987wc}. 
The mixed representation reads 
\be
\tilde{\Delta}(K) &=& \frac{1}{k_0^2-E_k^2} = -\int\limits_0^\beta d\tau 
e^{k_0\tau} \Delta_F(\tau,E_k),\label{Delta_tilde} \\
\Delta(K) &=& \frac{1}{k_0^2-k^2} =-\int\limits_0^\beta d\tau
e^{k_0\tau} \Delta_B(\tau,k),
\label{Delta_K}
\ee
with
\be
&&\hspace{-.5cm} \Delta_F(\tau,E_k)\nn
 &&=\frac{1}{2E_k}\left[\left(1-n_F^-(E_k)\right)e^{-E_k\tau} - n_F^+(E_k)e^{E_k\tau}
 \right],\label{Delta_F}\\
&&\hspace{-.5cm}\Delta_B(\tau,k) \nn
&&= 
\frac{1}{2k}\left[ n_B(k)e^{k\tau}+\left(1+n_B(k)\right)e^{-k\tau} \right]\!.\ 
\ee
Using Eqs.~(\ref{Delta_tilde}) and~(\ref{Delta_K}), the necessary Matsubara sum can be performed as
\begin{widetext}
\be
&&\hspace{-1cm}T\sum\limits_{k_0}\Delta(K)\tilde{\Delta}(P-K)\nn &=& \int\limits_0^\beta d\tau 
e^{p_0\tau} \Delta_B(\tau,k)\Delta_F(\tau,E_{pk})
=- \sum\limits_{s_1,s_2=\pm 1} \!\!
\frac{s_1s_2\left[1+n_B(s_1 k) - n_F^-(s_2E_{pk})\right]}{4 k E_{pk}
	\left(p_0-s_1k-s_2E_{pk}\right)}\!\hspace{.7cm}\nn
%%%%%%%%%%%%%%%%%%
&=&-\frac{1}{4 k E_{pk}}\Bigg[\frac{1+n_B(k) - n_F^-(E_{pk})}{p_0-k-E_{pk}}-\frac{1 + n_B(k) - n_F^+(E_{pk})}{p_0+k+E_{pk}}
-\frac{n_B(k) + n_F^+(E_{pk})}{p_0- k + E_{pk}} + \frac{n_B(k) + n_F^-(E_{pk})}{p_0 + k -E_{pk}}\Bigg]
\label{sumint_f1}
\ee
\end{widetext}
% % % % % % % % % % % % % % % % % % % % % 
Within the HTL approximation, the first two terms of Eq.~(\ref{sumint_f1}) can be neglected, and we can write the Matsubara sum as
\be
&&T\sum\limits_{k_0}\Delta(K)\tilde{\Delta}(P-K) \nn
&&\approx
\frac{1}{4kE_{pk}}\left[\frac{n_B(k)+n_F^+(E_{pk})}{p_0-k+E_{pk
}}-\frac{n_B(k)+n_F^-(E_{pk})}{p_0+k-E_{pk}}\right]\!. \hspace{.7cm}\label{tree1}
\ee
% % % % % % % % % % % 
\noindent
Following a similar procedure, we can calculate the second Matsubara sum as
\be
&&T\sum\limits_{k_0}\! k_0 \Delta(K)\tilde{\Delta}(P-K)\nn
&&\approx\frac{1}{4E_{pk}}\left[\frac{n_B(k) + n_F^+(E_{pk})}{p_0-k+E_{pk}}
	+\frac{n_B(k)+n_F^-(E_{pk})}{p_0-k+E_{pk}}\right]\!.\hspace{.5cm} \label{tree2}
\ee
Using the approximations in Eq.~(\ref{htl_approx}), 
 Eqs.~(\ref{tree1}) and (\ref{tree2}) can be simplified as
\be
T\sum\limits_{k_0}\tilde{\Delta}(K)\Delta(P-K)&\simeq  &
\frac{1}{4kE_k}\left[\frac{n_F^+(E_k)+n_B(k)}{p_0-k+E_k-v\bm{p
\cdot \hat{k}}}\right.\nn
&&\left.-\frac{n_F^-(E_k)+n_B(k)}{p_0+k-E_k+v\bm{p\cdot \hat{k}}}\right]\!\hspace{.7cm} 
\label{f_sum1}
\ee
%%%%%%%%%%%%%%%%%%%%%%%%%%%%%
and
\be
T\sum\limits_{k_0} k_0\tilde{\Delta}(K)\Delta(P-K)&\simeq  &2
\frac{1}{4E_k}\left[\frac{n_F^+(E_k)+n_B(k)}{p_0-k+E_k-v\bm{p
\cdot \hat{k}}}\right.\nn
&&\left.\hspace{-2.8cm}+\ \frac{n_F^-(E_k)+n_B(k)}{p_0+k-E_k+v\bm{p\cdot \hat{k}}}\right]\!.
\label{f_sum2}
\ee
%%%%%%%%%%

\noindent Using the Matsubara sums in Eqs.~(\ref{f_sum1}) and~(\ref{f_sum2}), the quark self-energy in Eq.~(\ref{quark_self}) becomes
\be
&&\hspace{-.5cm}\Sigma(P)\nn&=& 
2g^2C_F\!\int\!\frac{k^2dk}{2\pi^2}\!\int\!\frac{d\Omega}{4\pi}\Bigg[\frac{\gamma_0}{4E_k}\left\{\frac{n_F^+(E_k)+n_B(k)}{p_0-k+E_k - v\bm{p\cdot \hat{k}}}\right.\nn
&+&\left.\frac{n_F^-(E_k)+n_B(k)}{p_0+k-E_k+v\bm{p\cdot \hat{k}}}\right\}-\ \frac{\bm{\gamma\cdot \hat{k}}\ k-2m_f}{4kE_k}\nn
%%%%%%%%%%%%%%%%%%%%%%%%%%%%
&\times&\left\{\frac{n_F^+(E_k)+n_B(k)}{p_0-k+E_k-v\bm{p\cdot \hat{k}}}-\frac{n_F^-(E_k)+n_B(k)}{p_0+k-E_k+v\bm{p\cdot \hat{k}}}\right\}\!\Bigg]\nn
&=& 
\frac{g^2C_F}{4\pi^2}\!\!\int\!\frac{kdk}{E_k}\!\int\!\frac{d\Omega}{4\pi}\Bigg[\(n_F^+(E_k)+n_B(k)\)\frac{k\hat{\slashed{K}}+2m_f}{\tilde{P}^+\cdot\hat{K}}\nn
&&+\(n_F^-(E_k)+n_B(k)\)\frac{k\hat{\slashed{K}}-2m_f}{\tilde{P}^{-}\cdot\hat{K}}\Bigg],
\label{self_final}
\ee
with
\be
\tilde{P}^\pm\equiv \(\tilde{p}^\pm_{0},v\bm{p}\),\hspace{.1cm} \tilde{p}^\pm_{0}=p_0\pm(E_k-k), 
\hspace{.1cm} \hat{K}\equiv \(1,\bm{\hat{k}}\).%\hspace{1cm} \hat{K}'\equiv \(1,-\bm{\hat{k}}\)
\qquad
\ee
%%%%%%%%%%%%%%%%%%%%%%%%%%%%%
\noindent Using Eqs.~(\ref{abc}), (\ref{A0sm_def}), and (\ref{self_final}), we can write
\begin{widetext}
\be
A_0(p_0,p)&=&p_0-\frac{1}{4}\mbox{Tr}[\gamma_0\ \Sigma(P)]\nn
&=&p_0-2g^2C_F\int\frac{k^2dk}{2\pi^2}\int\frac{d\Omega}{4\pi}\frac{1}{4E_k}\left\{\frac{n_F^+(E_k)+n_B(k)}{\tilde{P}^+\cdot\hat{K}}+\frac{n_F^-(E_k)+n_B(k)}{\tilde{P}^-\cdot\hat{K}}\right\}\nn
%%%%%%%%%%%%%%%%%%%%%%%%%%%%%%%%%%%%%%%%%%%%%%%%
&=&p_0-\frac{g^2C_F}{8\pi^2}\int\! kdk\ \frac{1}{p}\left[\(n_F^+(E_k)+n_B(k)\)\log\frac{\tilde{p}_0^++vp}{\tilde{p}_0^+-vp}
+\(n_F^-(E_k)+n_B(k)\)\log\frac{\tilde{p}_0^-+vp}{\tilde{p}_0^--vp}\right],
\label{A0_final}\\
%%%%%%%%%%%%%%%%%%%%%%%%%%%%%%%%%%%%%%%%%%%%%%%%%%%
%%%%%%%%%         As          %%%%%%%%%%%%%%%%%%%%%
%%%%%%%%%%%%%%%%%%%%%%%%%%%%%%%%%%%%%%%%%%%%%%%%%%%
A_s(p_0,p) &=&p-\frac{1}{4}\mbox{Tr}[\bm{\gamma\cdot\hat{p}}\ \Sigma(P)]\nn
&=&p-2g^2C_F\int\frac{k^2dk}{2\pi^2}\int\frac{d\Omega}{4\pi}\frac{\bm{\hat{p}\cdot \hat{k}}}{4E_k}\left\{\frac{n_F^+(E_k)+n_B(k)}{\tilde{P}^+\cdot\hat{K}}+\frac{n_F^-(E_k)+n_B(k)}{\tilde{P}^-\cdot\hat{K}}\right\}\nn
%%%%%%%%%%%%%%%%%%%%%%%%%%%%%%%%%%%%%%%%%%%
&=& p + \frac{g^2C_F}{4\pi^2}\int\!kdk\ \frac{1}{p}\left[\(n_F^+(E_k)+n_B(k)\)\left\{1-\frac{\tilde{p}_0^+}{2vp}\log\frac{\tilde{p}_0^++vp}{\tilde{p}_0^--vp}\right\}\right.\nn
&&\hspace{3.75cm}+\left.\(n_F^-(E_k)+n_B(k)\)\left\{1-\frac{\tilde{p}_0^-}{2vp}\log\frac{\tilde{p}_0^-+vp}{\tilde{p}_0^+-vp}\right\}\right],
\label{As_final}\\ 
%%%%%%%%%%%%%%%%%%%%%%%%%%%%%%%%%%%%%%%%%%%%%%%%%%%
%%%%%%%%%         Am          %%%%%%%%%%%%%%%%%%%%%
%%%%%%%%%%%%%%%%%%%%%%%%%%%%%%%%%%%%%%%%%%%%%%%%%%%
A_m(p_0,p)
&=&m_f+\frac{1}{4}\mbox{Tr}[\Sigma(P)]\nn
%%%%%%%%%%%%%%%%%%%%%%%%%%%%%%%%%%%%%%%
&=&m_f+2g^2C_F\int\frac{k^2dk}{2\pi^2}\int\frac{d\Omega}{4\pi}\frac{m_f}{2kE_k}\left\{\frac{n_F^+(E_k)+n_B(k)}{\tilde{P}^+\cdot\hat{K}}-\frac{n_F^-(E_k)+n_B(k)}{\tilde{P}^-\cdot\hat{K}}\right\}\nn
%%%%%%%%%%%%%%%%%%%%%%%%%%%%%%%%%%%%%%%
&=&m_f+ \frac{g^2C_F}{4\pi^2}\int\! dk\ \frac{m_f}{p}\left[\(n_F^+(E_k)+n_B(k)\)\log\frac{\tilde{p}_0^++vp}{\tilde{p}_0^+-vp}-\(n_F^-(E_k)+n_B(k)\)\log\frac{\tilde{p}_0^-+vp}{\tilde{p}_0^--vp}\right].
\label{Am_final}
\ee
\end{widetext}
% % % % % % % % % % % % % % % % % 
\noindent 
Now, we can construct $D_\pm(p_0,p)$ from Eq.~(\ref{Dpm_def}) using Eqs.~(\ref{A0_final})-(\ref{Am_final}). We can get the quark dispersion relation at a finite quark mass by solving $D_\pm(p_0,p)=0$.
\begin{figure}[tbh!]
	\includegraphics[width=6.6cm]{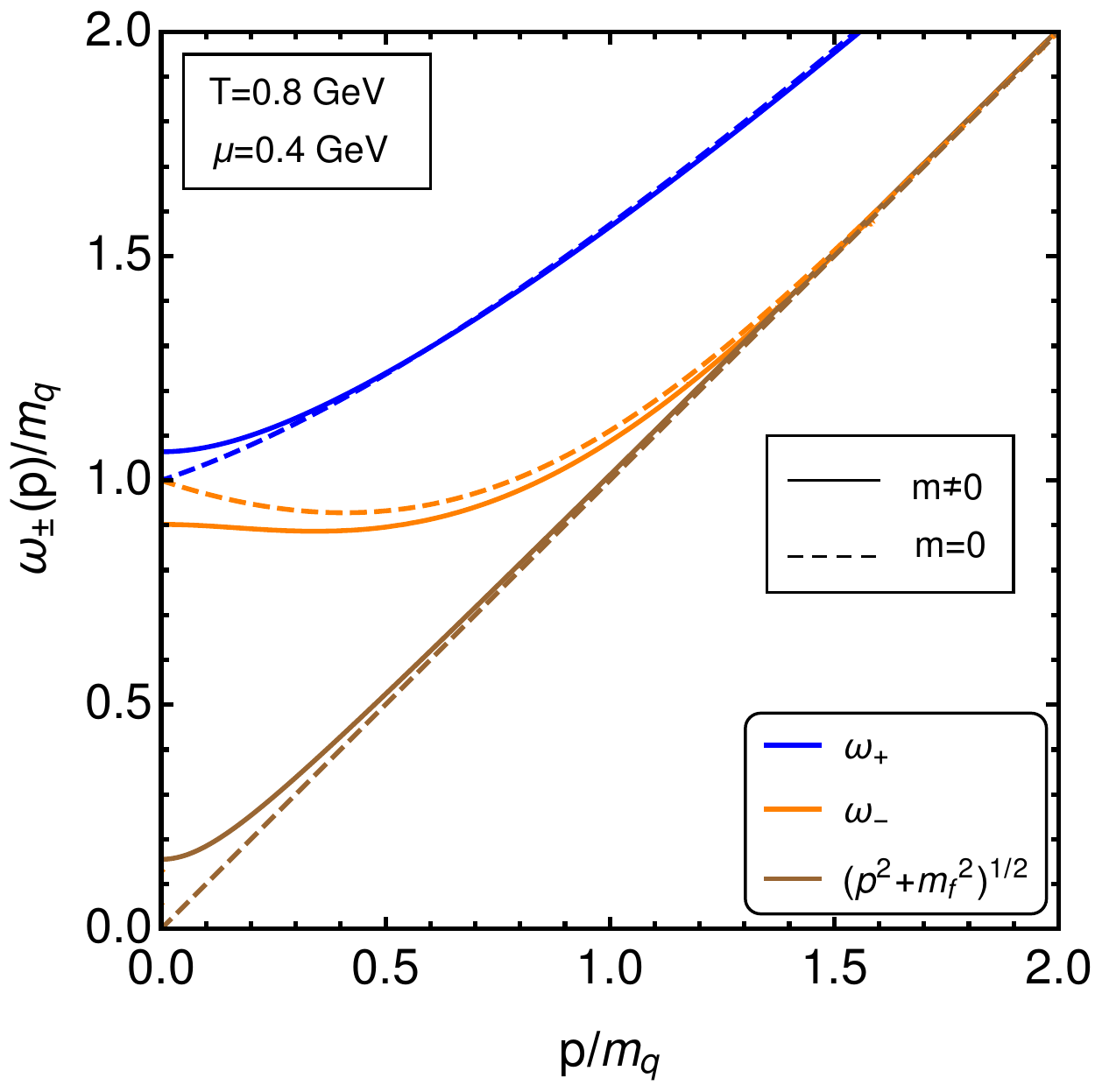}
	\caption{Dispersion plots for the strange quark. The solid lines represent dispersion relations for the massive quark, whereas the dashed lines represent corresponding massless modes.}
	\label{fig:squark_disp}
\end{figure} 
%%%%%%%%%%%% 

\noindent In Fig.~\ref{fig:squark_disp}, we plot the dispersion relation for the massive strange quark. The solid lines represent the dispersion relation for the massive quark, whereas dashed lines represent corresponding massless modes.
%%%%%%%%%%%%%%%%%%%%%%%%%%%%%
In Fig.~\ref{fig:squark_disp} and also the remaining part of the paper, we use the scale-dependent strange quark mass~\cite{Graf:2015tda} as
\be
m_s(\Lambda)=\hat{m}_s\(\frac{\alpha_s}{\pi}\)^{4/9}
%\left[1+0.895062\frac{\alpha_s}{\pi}\right]
,\label{ms}
\ee
where running coupling $\alpha_s=g^2/4\pi$ and we use one-loop running coupling for the numerical computation as
\be
\alpha_{s}&=&\frac{12\pi}{11N_c - 2N_f}\frac{1}{ 
	\ln(\Lambda^2/\Lambda_{\overline{\rm MS}}^2)}.
\ee
The scale $\Lambda_{\overline{\rm MS}}$ is fixed by requiring $\alpha_{s}=0.326$ at $\Lambda = 1.5$ GeV~\cite{Bazavov:2012ka} whereas the invariant mass $\hat{m}_s$ is fixed by requiring $m_s=100$ MeV at $\Lambda = 2$ GeV~\cite{Patrignani:2016xqp}; one obtains $\Lambda_{\overline{\rm MS}}=176 $ MeV and
$\hat{m}_s=290$ MeV. For all dispersion plots, we take the renormalization scale as $\Lambda=2\pi\sqrt{T^2+(\mu/\pi)^2}$.

Note that unlike the vanishing quark masses case, integration over loop momentum $k$ in Eqs.~(\ref{A0_final})\textendash(\ref{Am_final}) cannot be done analytically. As a consequence, one may use the expressions in Eqs.~(\ref{A0_final})\textendash(\ref{Am_final}) only to calculate some quantities in one-loop order. For a higher-loop computation, such as three-loop thermodynamics, one may not be able to proceed with general expressions of $A_0(p_0,p), A_s(p_0,p),$ and $A_m(p_0,p)$. Keeping that in mind, we can simplify the expressions of $A_0(p_0,p), A_s(p_0,p),$ and $A_m(p_0,p)$ at the high temperature limit $(m_f\ll T)$ as
% % % % % 
\be
A_0(p_0,p)&\approx& p_0 -\frac{m_q^2}{2p}\log\frac{p_0+p}{p_0-p}, \nn
A_s(p_0,p)&\approx& p_0 +\frac{m_q^2}{p}\left[1-\frac{p_0}{2p}\log\frac{p_0+p}{p_0-p}\right], \nn
A_m(p_0,p)&\approx& m_f,
\label{A0AsAm_approx}
\ee
where $m_q$ is the quark thermal mass and can be written as
\be
m_q^2=\frac{g^2T^2C_F}{8}\(1+\frac{\mu_f^2}{\pi^2T^2}\).
\ee
Equation~(\ref{A0AsAm_approx}) can also be obtained from Eq.~(\ref{med_prop_inv_def}) considering massless one-loop quark self-energy (normal HTL results) and massive free quark propagator $S_0(P) = i/\left(\slashed{P}-m_f\right)$. 
%%%%%%%

\begin{figure}[tbh!]
 \includegraphics[width=6.6cm]{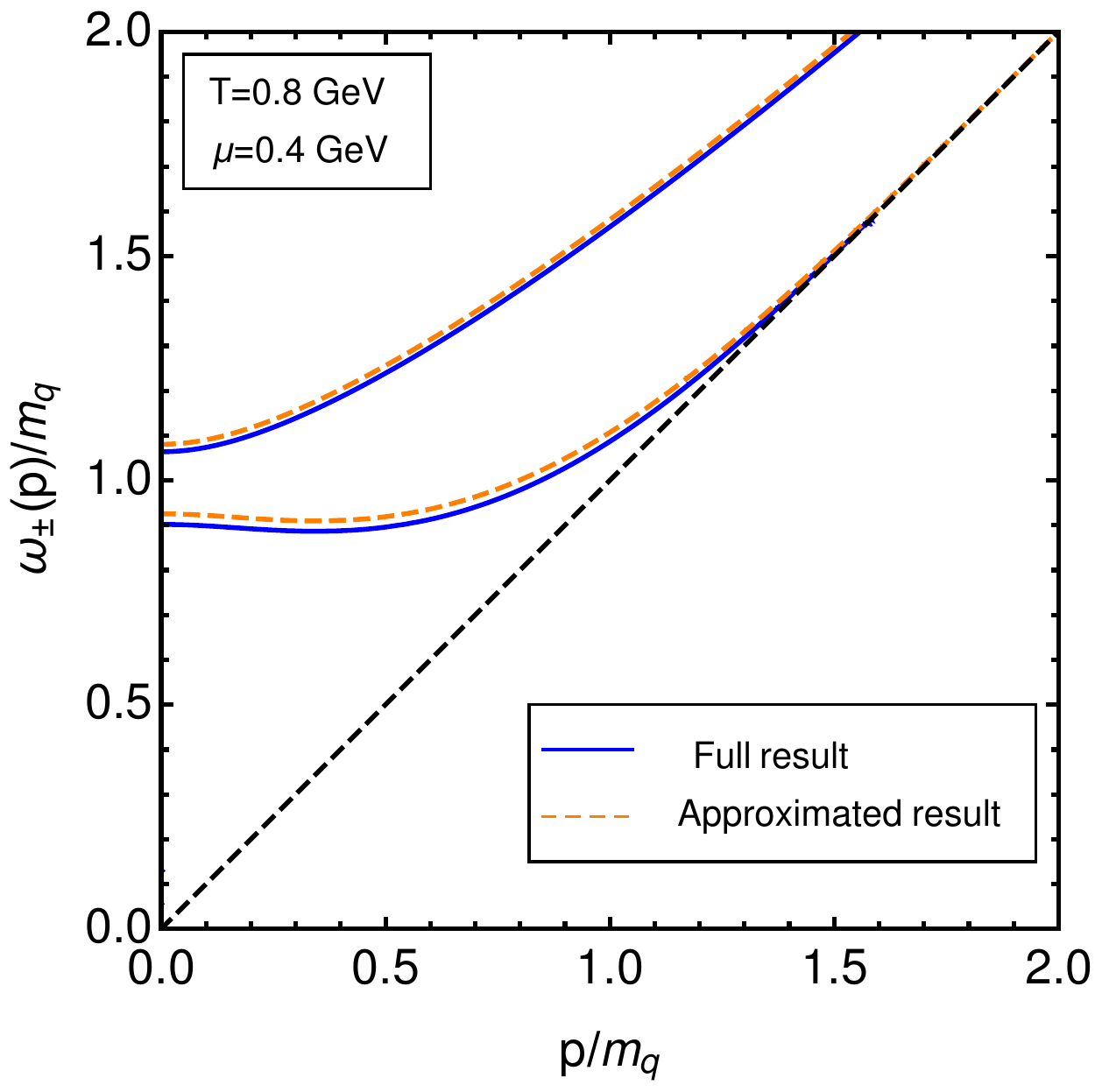}
	\caption{Dispersion plots for the strange quark. Dispersion relations from the full quark propagator are compared with the approximated one in Eq.~(\ref{A0AsAm_approx}).}
	\label{fig:squark_disp_approx}
\end{figure}
\noindent In Fig.~\ref{fig:squark_disp_approx}, we compare the full dispersion relation of the massive s quark along with the dispersion relations that have been obtained considering  $m_f\ll T$. From Fig.~\ref{fig:squark_disp_approx}, it is clear that the approximated results in Eq.~(\ref{A0AsAm_approx}) are a good approximation at high temperature.

\noindent For completeness, we discuss three- and four-point quark-gluon vertices without a high temperature limit $(m_f\ll T)$ in the next section.

\section{Three- and four-point quark-gluon vertices}
\label{sec:3vertex}
\begin{figure}[h]
		\includegraphics[width=8cm]{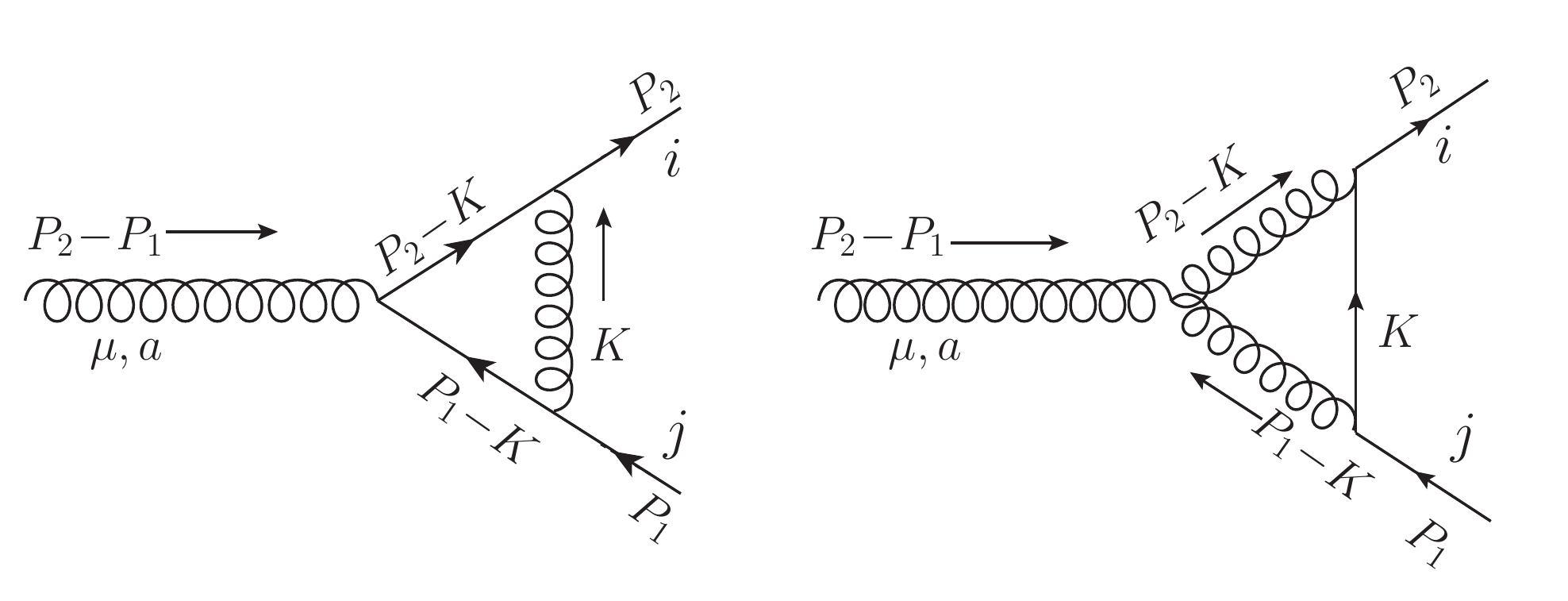}
	\caption{Three-point quark-gluon vertex}
\label{fig:qqg_vertex}
\end{figure}

\noindent The one-loop Feynman diagrams for the quark-gluon vertex are illustrated in Fig.~\ref{fig:qqg_vertex}.

\noindent
The one-loop vertex correction from the two diagrams in the Feynman gauge can be written as
\begin{widetext}
\be
ig\ \delta\Gamma^{1,a}_{\mu,ij}(P_1,P_2)&=&\int\frac{d^4K}{(2\pi)^4} \(ig\gamma^\alpha T^b_{il'}\) S_0^{l'l}(P_2-K)\(ig\gamma_\mu T^a_{lm}\)  S_0^{mm'}(P_1-K) \(ig\gamma^\beta T^{b'}_{m'j}\) \frac{-ig_{\alpha\beta}\delta^{bb'}}{K^2}\nn
&=&-i(ig)^3(T^bT^aT^b)_{ij}\int\frac{d^4K}{(2\pi)^4}\ \gamma^\alpha S_0(P_2-K) \gamma_\mu  S_0(P_1-K) \gamma_\alpha \frac{1}{K^2}\nn
&=&-g^3\(C_F-\frac{C_A}{2}\)T^a_{ij}\int\frac{d^4K}{(2\pi)^4}\ \gamma^\alpha S_0(P_2-K) \gamma_\mu  S_0(P_1-K) \gamma_\alpha \frac{1}{K^2}
\ee
and

\be
ig\ \delta\Gamma^{2,a}_{\mu,ij}(P_1,P_2)&=&-igf_{abc}\int\frac{d^4K}{(2\pi)^4}\ V_{\mu\alpha\beta}\ \frac{-ig^{\alpha\alpha'}\delta^{bb'}}{(P_2-K)^2} \(ig\gamma_{\alpha'} T^{b'}_{il}\) S_0^{ll'}(K)\(ig\gamma_{\beta'} T^{c'}_{lj}\)   \frac{-ig^{\beta\beta'}\delta^{cc'}}{(P_1-K)^2}\nn
&=&igf_{abc}(ig)^2(T^bT^c)_{ij}\int\frac{d^4K}{(2\pi)^4}\ V_{\mu\alpha\beta}\ \gamma^\alpha S_0(K) \gamma^\beta  \frac{1}{(P_1-K)^2(P_2-K)^2}\nn
&=&g^3\frac{C_A}{2}T^a_{ij} \int\frac{d^4K}{(2\pi)^4}\ V_{\mu\alpha\beta}\  \gamma^\alpha S_0(K) \gamma^\beta  \frac{1}{(P_1-K)^2(P_2-K)^2},
\ee

where 
\be
V_{\mu\alpha\beta} = g_{\mu\alpha}\(2P_2-2P_1+K\)_\beta + g_{\alpha\beta}\(P_1-P_2 - 2K\)_\mu + g_{\beta\mu}\(P_1-P_2+K\)_\alpha.
\ee
\noindent We can define one-loop vertex correction $\delta\Gamma^{n,a}_{\mu,ij}(P_1,P_2)$ in terms of vertex correction function $\delta\Gamma^{n,a}(P_1,P_2)$ as
\be
\delta\Gamma^{n,a}_{\mu,ij}(P_1,P_2)=\delta\Gamma^n_\mu(P_1,P_2)\ T^{a}_{ij}.
\ee

\noindent
The one-loop vertex correction function $\delta\Gamma^{1}_\mu(P_1,P_2)$ from the first diagram is
\be
\delta\Gamma^1_\mu(P_1,P_2)&=&ig^2\(C_F-\frac{C_A}{2}\)\!\int\!\!\frac{d^4K}{(2\pi)^4}\gamma^\alpha S_0(P_2-K)\gamma_\mu  S_0(P_1-K)\gamma_\alpha \frac{1}{K^2}\nn
&=&-ig^2\(C_F-\frac{C_A}{2}\)\!\int\!\!\frac{d^4K}{(2\pi)^4}\ \gamma^\alpha \(\slashed{K}-m_f\)\gamma_\mu  \(\slashed{K}-m_f\)\gamma_\alpha \tilde{\Delta}(P_1-K)\tilde{\Delta}(P_2-K)\Delta(K)\nn
&=&-2ig^2\(C_F-\frac{C_A}{2}\)\!\int\!\!\frac{d^4K}{(2\pi)^4} \Big[\!\(K^2-m^2\)\gamma_\mu-2K_\mu(\slashed{K}+2m_f)\Big]
\tilde{\Delta}(P_1-K)\tilde{\Delta}(P_2-K)\Delta(K).
\label{Gamma_1_def}
\ee

\noindent The first term within the square brackets in Eq.~(\ref{Gamma_1_def}) is nonleading in temperature and can be neglected using the HTL approximation. The remaining contribution can be written as
%%%%%%%
\be
\hspace{-1cm}\delta\Gamma^1_\mu(P_1,P_2)&\approx&4ig^2\(C_F-C_A/2\)\int\frac{d^4K}{(2\pi)^4}K_\mu(\slashed{K}+2m_f)\tilde{\Delta}(P_1-K)\tilde{\Delta}(P_2-K)\Delta(K)\nn
&=&-4g^2\(C_F-C_A/2\)T\sum_n\int\frac{d^3k}{(2\pi)^3}K_\mu(\slashed{K}+2m_f)\tilde{\Delta}(P_1-K)\tilde{\Delta}(P_2-K)\Delta(K).
\label{Gamma_1_def2}
\ee
 %\end{widetext}
We need to calculate the following Matsubara sums:
\be
X_0&=&T\sum_n \Delta(K) \tilde{\Delta}(P_1-K)\tilde{\Delta}(P_2-K),\label{X0_def}\\
X_1&=&T\sum_n k_0\Delta(K)\tilde{\Delta}(P_1-K)\tilde{\Delta}(P_2-K),\label{X1_def}\\
X_2&=&T\sum_n k_0^2\Delta(K)\tilde{\Delta}(P_1-K)\tilde{\Delta}(P_2-K).\label{X2_def}
\ee
%\begin{widetext}
\noindent Now,
\be
X_0&=&T\sum_n \Delta(K) \tilde{\Delta}(P_1-K)\tilde{\Delta}(P_2-K)\nn
&=&T\sum_{n,s,s_1,s_2}\frac{ss_1s_2}{8EE_1E_2}\frac{1}{(k_0-sE)((p_{10}-k_0)-s_1E_1)((p_{20}-k_0)-s_2E_2)}\nn
&=& T\sum_{n,s,s_1,s_2}\frac{ss_1s_2}{8EE_1E_2}\frac{1}{(p_{10}-p_{20})-s_1E_1+s_2E_2}\left[\frac{1}{(k_0-sE)((p_{20}-k_0)-s_2E_2)}-\frac{1}{(k_0-sE)((p_{10}-k_0)-s_1E_1)}\right]\nn
&=& -\sum_{s,s_1,s_2}\frac{ss_1s_2}{8EE_1E_2}\frac{1}{(p_{10}-p_{20})-s_1E_1+s_2E_2}\left[\frac{1+n_B(sE)-n_F^-(s_2E_2)}{p_{20}-sE-s_2E_2}-\frac{1+n_B(sE)-n_F^-(s_1E_1)}{p_{10}-sE-s_1E_1}\right]\nn
%%%%%%%%%%%%%%%%%%%%%%%%%%%
&\approx&-\frac{1}{8EE_1E_2}\Bigg[\frac{1}{(p_{10}-p_{20})+E_1-E_2}\Bigg\{\frac{n_B(E)+n_F^+(E_2)}{p_{20}-E+E_2}-\frac{n_B(E)+n_F^+(E_1)}{p_{10}-E+E_1}\Bigg\}\nn
&&+\frac{1}{(p_{10}-p_{20})-E_1+E_2}\Bigg\{\frac{n_B(E)+n_F^-(E_2)}{p_{20}+E-E_2}-\frac{n_B(E)+n_F^-(E_1)}{p_{10}+E-E_1}\Bigg\}\Bigg],
\label{X0_ini}
\ee
where $E=k,\ E_1=\sqrt{|\bm{p_1-k}|^2+m_f^2},\ $ and $E_2=\sqrt{|\bm{p_2-k}|^2+m_f^2}$.

\noindent Applying the HTL approximation, we can write Eq.~(\ref{X0_ini}) as
%%%%%%%%%%%%%%%%%%%%%%%%%%%
\be
X_0
&\approx&-\frac{1}{8kE_k^2}\Bigg[\frac{n_B(k)+n_F^+(E_k)}{(p_{10}-k+E_k-v\bm{p_1\cdot \hat{k}})(p_{20}-k+E_k-v\bm{p_2\cdot \hat{k}})}
+\frac{n_B(k)+n_F^-(E_k)}{(p_{10}+k-E_k+v\bm{p_1\cdot \hat{k}})(p_{20}+k-E_k+v\bm{p_2\cdot \hat{k}})}\Bigg]\nn
%%%%%%%%%%%%%%%%%%%%%%%%%%%
 &=&-\frac{1}{8kE_k^2}\Bigg[\frac{n_B(k)+n_F^+(E_k)}{\(\tilde{P}^+_{1}\cdot \hat{K}\)\(\tilde{P}^+_{2}\cdot \hat{K}\)}
 +\frac{n_B(k)+n_F^-(E_k)}{\(\tilde{P}^-_{1}\cdot \hat{K}'\)\(\tilde{P}^-_{2}\cdot \hat{K}'\)}\Bigg],
 \label{X0_final}
%%%%%%%%%%%%%%%%%%%%%%%%%%%
\ee
where $\hat{K}'=\(1,-\hat{\bm{k}}\)$.
\end{widetext}
The next Matsubara frequency $X_1$ from Eq.~(\ref{X2_def}) can be evaluated using $k_0\rightarrow sE$ as
\be
X_1&=&-\frac{1}{8E_k^2}\Bigg[\frac{n_B(k)+n_F^+(E_k)}{\(\tilde{P}^+_{1}\cdot \hat{K}\)\(\tilde{P}^+_{2}\cdot \hat{K}\)}\nn
&&
 -\ \frac{n_B(k)+n_F^-(E_k)}{\(\tilde{P}^-_{1}\cdot \hat{K}'\)\(\tilde{P}^-_{2}\cdot \hat{K}'\)}\Bigg].
  \label{X1_final}
\ee
Similarly,
\be
X_2\simeq k^2X_0
&=& -\frac{k}{8E_k^2}\Bigg[\frac{n_B(k)+n_F^+(E_k)}{\(\tilde{P}^+_{1}\cdot \hat{K}\)\(\tilde{P}^+_{2}\cdot \hat{K}\)}\nn
&&
 +\ \frac{n_B(k)+n_F^-(E_k)}{\(\tilde{P}^-_{1}\cdot \hat{K}'\)\(\tilde{P}^-_{2}\cdot \hat{K}'\)}\Bigg].
  \label{X2_final}
\ee
% % % % % % % % % % % % % % 
\noindent Combining all these Matsubara sums from Eqs.~(\ref{X0_final})-~(\ref{X2_final}), we can write the first diagram of the three-point quark-gluon vertex from Eq.~(\ref{Gamma_1_def2}) as
\be
&&\hspace{-1cm}\delta\Gamma^1_\mu(P_1,P_2)\nn&=&4g^2\(C_F-C_A/2\)\int\frac{d^3k}{(2\pi)^3}\frac{1}{8E_k^2}\nn
&\times&\Bigg[\(n_B(k)+n_F^+(E_k)\)\frac{\hat{K}_\mu\(k\hat{\slashed K}+2m_f\)}{\(\tilde{P}^+_{1}\cdot \hat{K}\)\(\tilde{P}^+_{2}\cdot \hat{K}\)}\nn
&+&\(n_B(k)+n_F^-(E_k)\)\frac{\hat{K}_\mu\(k\hat{\slashed K}-2m_f\)}{\(\tilde{P}^-_{1}\cdot \hat{K}\)\(\tilde{P}^-_{2}\cdot \hat{K}\)}\Bigg],
\ee
%%%%%%%%%%%
Following the same procedure, the second diagram of the three-point quark-gluon vertex can be written as
\be
&&\hspace{-1cm}\delta\Gamma^2_\mu(P_1,P_2)\nn&\simeq&4g^2C_A/2\int\frac{d^3k}{(2\pi)^3}\frac{1}{8E_k^2}\nn
 &\times&\Bigg[\(n_B(k)+n_F^+(E_k)\)\frac{\hat{K}_\mu\(k\hat{\slashed K}+2m_f\)}{\(\tilde{P}^+_{1}\cdot \hat{K}\)\(\tilde{P}^+_{2}\cdot \hat{K}\)}\nn
&&+\(n_B(k)+n_F^-(E_k)\)\frac{\hat{K}_\mu\(k\hat{\slashed K}-2m_f\)}{\(\tilde{P}^-_{1}\cdot \hat{K}\)\(\tilde{P}^-_{2}\cdot \hat{K}\)}\Bigg]
\ee
% % % % % % % % % 
\noindent So, the total three-point quark-gluon vertex becomes
\be
&&\hspace{-1cm}\Gamma_\mu(P_1,P_2)\nn
&=&\gamma_\mu+\delta\Gamma^1_\mu(P_1,P_2)+\delta\Gamma^2_\mu(P_1,P_2)\nn
&=&\gamma_\mu + \frac{g^2C_F}{4\pi^2}\int \frac{k^2dk}{E_k^2}\int\frac{d\Omega}{4\pi}\nn
&\times&\Bigg[\(n_B(k)+n_F^+(E_k)\)\frac{\hat{K}_\mu\(k\hat{\slashed K}+2m_f\)}{\(\tilde{P}^+_{1}\cdot \hat{K}\)\(\tilde{P}^+_{2}\cdot \hat{K}\)}\nn
&&+\(n_B(k)+n_F^-(E_k)\)\frac{\hat{K}_\mu\(k\hat{\slashed K} - 2m_f\)}{\(\tilde{P}^-_{1}\cdot \hat{K}\)\(\tilde{P}^-_{2}\cdot \hat{K}\)}\Bigg].
\label{3vertex_final}
\ee
\begin{widetext}
	
\noindent Now, the Ward identity (WI) of the three-point
quark-gluon vertex can be checked as
\be
&&\hspace{-2cm}(P_1-P_2)^\mu\ \Gamma_\mu(P_1,P_2)\nn
&=&\slashed{P}_1-\slashed{P}_2 + \frac{g^2C_F}{4\pi^2}\int \frac{k^2dk}{E_k^2}\int\frac{d\Omega}{4\pi} \Bigg[\(n_B(k)+n_F^+(E_k)\)\frac{\(P_1\cdot\hat{K}-P_2\cdot\hat{K}\)\(k\hat{\slashed K}+2m_f\)}{\(\tilde{P}^+_{1}\cdot \hat{K}\)\(\tilde{P}^+_{2}\cdot \hat{K}\)}\nn
&&+\(n_B(k)+n_F^-(E_k)\)\frac{\(P_1\cdot\hat{K}-P_2\cdot\hat{K}\)\(k\hat{\slashed K} - 2m_f\)}{\(\tilde{P}^-_{1}\cdot \hat{K}\)\(\tilde{P}^-_{2}\cdot \hat{K}\)}\Bigg]\nn
%%%%%%%%%%%%%%%%%%%%%%%%%%%%%%%%%%%%%%%%%%%%%%%
&\simeq&\slashed{P}_1-\slashed{P}_2 + \frac{g^2C_F}{4\pi^2}\int \frac{kdk}{E_k}\int\frac{d\Omega}{4\pi} \Bigg[\(n_B(k)+n_F^+(E_k)\)\frac{\(\tilde{P}^+_{1}\cdot\hat{K}-\tilde{P}^+_{2}\cdot\hat{K}\)\(k\hat{\slashed K}+2m_f\)}{\(\tilde{P}^+_{1}\cdot \hat{K}\)\(\tilde{P}^+_{2}\cdot \hat{K}\)}\nn
&&+\(n_B(k)+n_F^-(E_k)\)\frac{\(\tilde{P}^-_{1}\cdot\hat{K}-\tilde{P}^-_{2}\cdot\hat{K}\)\(k\hat{\slashed K} - 2m_f\)}{\(\tilde{P}^-_{1}\cdot \hat{K}\)\(\tilde{P}^-_{2}\cdot \hat{K}\)}\Bigg]\nn
%%%%%%%%%%%%%%%%%%%%%%%%%%%%%%%%%%%%%%%%%%%%%%%
&=&\slashed{P}_1-m_f - \frac{g^2C_F}{4\pi^2}\int\! \frac{kdk}{E_k}\frac{d\Omega}{4\pi} \Bigg[\(n_B(k)+n_F^+(E_k)\)\frac{k\hat{\slashed K}+2m_f}{\tilde{P}^+_{1}\cdot \hat{K}}+\(n_B(k)+n_F^-(E_k)\)\frac{k\hat{\slashed K} - 2m_f}{\tilde{P}^-_{1}\cdot \hat{K}}\Bigg]\nn
&-&\slashed{P}_2+m_f + \frac{g^2C_F}{4\pi^2}\int\! \frac{kdk}{E_k}\frac{d\Omega}{4\pi} \Bigg[\(n_B(k)+n_F^+(E_k)\)\frac{k\hat{\slashed K}+2m_f}{\tilde{P}^+_{1}\cdot \hat{K}}+\(n_B(k)+n_F^-(E_k)\)\frac{k\hat{\slashed K} - 2m_f}{\tilde{P}^-_{1}\cdot \hat{K}}\Bigg]\nn
&=&iS^{-1}(P_1)-iS^{-1}(P_2).
\label{WI}
\ee
Equation~(\ref{WI}) suggests that the three-point vertex satisfies the WI with the quark propagator. Note that the WI in Eq.~(\ref{WI}) is not exact; we consider a small quark mass approximation to get Eq.~(\ref{WI}). 
Additionally, without doing an exact calculation, one can also predict the form of the three-point vertex~[Eq.~(\ref{3vertex_final})] using the WI.

In a similar manner, one can also predict the form of the four-point quark-gluon vertex from the WI
\be
Q_1^\nu\ \Gamma_{\mu\nu}(P_1,P_2;Q_1)&=&  \Gamma_\mu(P_1,P_2-Q_1)-\Gamma_\mu(P_2,P_1+Q_1),
\ee
without calculating the Feynman diagram, as
\be
\Gamma_{\mu\nu}(P_1,P_2;Q_1)
&=& \frac{g^2C_F}{4\pi^2}\int \frac{k^2dk}{E_k^2}\int\frac{d\Omega}{4\pi}\nn
&\times&\Bigg[\(n_B(k)+n_F^+(E_k)\)\frac{\hat{K}_\mu\hat{K}_\nu\(k\hat{\slashed K}+2m_f\)}{\left[\(\tilde{P}^+_{1}+Q_1\)\cdot \hat{K}\right]\left[\(\tilde{P}^+_{2}-Q_1\)\cdot \hat{K}\right]}\left(\frac{1}{\tilde{P}^+_{1}\cdot \hat{K}}+\frac{1}{\tilde{P}^+_{2}\cdot \hat{K}}\right)\nn
&&+\(n_B(k)+n_F^-(E_k)\)\frac{\hat{K}_\mu\(k\hat{\slashed K} - 2m_f\)}{\left[\(\tilde{P}^-_{1}+Q_1\)\cdot \hat{K}\right]\left[\(\tilde{P}^-_{2}-Q_1\)\cdot \hat{K}\right]}\left(\frac{1}{\tilde{P}^+_{1}\cdot \hat{K}}+\frac{1}{\tilde{P}^+_{2}\cdot \hat{K}}\right)\Bigg].
\ee

In the next section, we study the effect of finite quark masses to thermodynamical quantities. To do that, we use finite mass modified effective quark propagator to calculate second-order quark and baryon number susceptibilities.
\section{Second-order susceptibilities}
\label{sec:qns}
\noindent The second-order QNS is the response of the quark number density with an infinitesimal
change in the quark chemical potential. Mathematically, the diagonal QNS can be written in terms of the number density as
\be
\chi_{ff}(T)=\left.\frac{\partial^2\mathcal{P}(T,\bm{\mu},\bm{m})}{\partial\mu_f^2}\right|_{\mu_f\rightarrow 0}=\left.\frac{\partial\mathcal{N}_f(T,\mu_f,m_f)}{\partial\mu_f}\right|_{\mu_f\rightarrow 0},
\ee
where $\mathcal{N}_f(T,\mu_f,m_f)$ represents the quark number density for the quark flavor $f$, $\mathcal{P}(T,\bm{\mu},\bm{m})$ represents pressure of the system, $\bm{m}=\{m_u,m_d,m_s,\cdots\}$, and $\bm{\mu}=\{\mu_u,\mu_d,\mu_s,\cdots\}$.
% % % % 

\noindent There are various ways to calculate the QNS within the HTL resummation framework, such as (i) calculating pressure by expanding it at small Debye and thermal quark masses and then taking the double derivative of that pressure with respect to the chemical potential~\cite{Haque:2013qta}, (ii) using the fluctuation-dissipation theorem~\cite{Chakraborty:2001kx}, and (iii) using two-loop approximately self-consistent $\Phi$-derivable HTL resummation~\cite{Blaizot:2001vr}. In the second and third methods, one uses quark quasiparticle poles $(\omega_\pm)$ to calculate the susceptibilities. As we have discussed the quark quasiparticle poles in the presence of finite quark masses in detail in Sec.~\ref{sec:quark_prop}, we use those results to calculate susceptibilities using method (iii).
 
 Now, the quark number density $\mathcal{N}_f(T,\mu_f,m_f)$ in terms of the dressed fermion propagators~\cite{Blaizot:2000fc,Blaizot:2001vr} in two-loop approximately self-consistent $\Phi$-derivable HTL resummation is given by
\be
&&\hspace{-.8cm}\mathcal{N}_f(T,\mu_f,m_f)\nn
&=&-2N_c\!\int\!\frac{d^4P}{(2\pi)^4}\frac{\partial n_F^+}{\partial \mu_f} 
\mbox{Tr}\Big[\mbox{Im}\log\(\gamma_0iS^{-1}(P)\)- 
\mbox{Im}\(\gamma_0\Sigma(P)\)\mbox{Re}\(iS(P)\gamma_0\)\Big]\nn
% % % % % % % % % % % % % % 
&=&-4N_c\!\int\!\!\frac{d^4P}{(2\pi)^4}\frac{\partial n_F^+}{\partial\mu_f} \Bigg\{\mbox{Im}\log D_+(p_0,p) + \mbox{Im}\log D_-(p_0,p)  + 
\mbox{Im}\left[p_0-A_0(p_0,p)\right]\mbox{Re}\left[\frac{1}{D_+(p_0,p)}+\frac{1}{D_-(p_0,p)}\right]\nn
%%%%%%%%%%%%%%%%%
&&-\mbox{Im}\left[p-A_s(p_0,p)\right]\mbox{Re}\left[\frac{A_s(p_0,p)}{D_+(p_0,p)}-\frac{A_s(p_0,p)}{D_-(p_0,p)}\right]+\mbox{Im}\left[m_f-A_m(p_0,p)\right]\mbox{Re}\left[\frac{A_m(p_0,p)}{D_+(p_0,p)}-\frac{A_m(p_0,p)}{D_-(p_0,p)}\right]\Bigg\}.
\ee
So, the second-order diagonal QNS can be obtained as
\be
\chi_{ff}(T)
&=&-\lim_{\mu_f\rightarrow0}4N_c\!\int\!\!\frac{d^4P}{(2\pi)^4}\frac{\partial^2 n_F^+}{\partial\mu_f^2}\nn
&\hspace{-1cm}\times&\hspace{-.5cm} \Bigg\{\mbox{Im}\log D_+(p_0,p) + \mbox{Im}\log D_-(p_0,p)  + 
\mbox{Im}\left[p_0-A_0(p_0,p)\right]\mbox{Re}\left[\frac{1}{D_+(p_0,p)}+\frac{1}{D_-(p_0,p)}\right]\nn
%%%%%%%%%%%%%%%%%
&\hspace{-1cm}-&\hspace{-.5cm}\mbox{Im}\left[p-A_s(p_0,p)\right]\mbox{Re}\left[\frac{A_s(p_0,p)}{D_+(p_0,p)}-\frac{A_s(p_0,p)}{D_-(p_0,p)}\right]+\mbox{Im}\left[m_f-A_m(p_0,p)\right]\mbox{Re}\left[\frac{A_m(p_0,p)}{D_+(p_0,p)}-\frac{A_m(p_0,p)}{D_-(p_0,p)}\right]\!\Bigg\}.
\label{chi2_def}
\ee
%%%%%%%
Note that the $\mu_f$ derivative is applied only to the explicit chemical potential dependence as discussed in Refs.~\cite{Blaizot:2000fc,Blaizot:2001vr}.
%%%%%%%
Now, the first two terms within curly brackets in Eq.~(\ref{chi2_def}) give a quasiparticle contribution to the QNS and can be obtained as
%%%
\be
\mathcal{\chi}_{ff}^{QP}(T,m_f)&=&\lim_{\mu_f\rightarrow0}N_c\sum_f^{N_f}\int\frac{p^2dp}{\pi^2}\frac{\partial^2 }{\partial \mu_f^2}
\Bigg[T\log\(1+e^{-\left[\omega_+-\mu_f\right]/T}\)+\ T\log\frac{1+e^{-\left[\omega_--\mu_f\right]/T}}{1+e^{-\left[k-\mu_f\right]/T}}
\nn
&+&  T\log\(1+e^{-\left[\omega_++\mu_f\right]/T}\) + T\log\frac{1+e^{-\left[\omega_-+\mu_f\right]/T}}{1+e^{-\left[k + \mu_f\right]/T}} \Bigg]\nn
&=&2N_c\beta\int\frac{p^2dp}{\pi^2}\Bigg[\frac{e^{\beta\omega_+}}{\(e^{\beta\omega_+}+1\)^2}
+\frac{e^{\beta\omega_-}}{\(e^{\beta\omega_-}+1\)^2}-\frac{e^{\beta k}}{\(e^{\beta k}+1\)^2} \Bigg].
\ee
%\end{widetext}
We identify the remaining contribution as the Landau-damping term that appears from the imaginary parts of the logarithmic terms within $A_0(p_0,p),A_s(p_0,p),$ and $A_m(p_0,p)$ and is obtained as
	\be
	\chi_{ff}^{LD}(T,m_f)
	&=&-N_c\beta^2 \int\frac{p^2dp}{\pi^3}\int\limits_{-\infty}^{\infty}d\omega\left[\frac{e^{\omega/T}\(e^{\omega/T}-1\)}{\(e^{\omega/T}+1\)^3}\right] \nn
	&\times&\Bigg\{\mbox{arg}D_+(p_0,p)+\mbox{arg}D_-(p_0,p)
	+\mbox{Im}[p_0-A_0(p_0,p)]\,\mbox{Re}\left[\frac{1}{D_+(p_0,p)}+\frac{1}{D_-(p_0,p)} \right] \nn
	&+&\mbox{Im}[A_s(p_0,p)-p]\mbox{Re}\left[\frac{A_s(p_0,p)}{\sqrt{A_s^2(p_0,p)+A_m^2(p_0,p)}}\left(\frac{1}{D_+(p_0,p)} - \frac{1}{D_-(p_0,p)} \right)\right]\nn
	&-&\left.\mbox{Im}[A_m(p_0,p)-m_f]\mbox{Re}\left[\frac{A_m(p_0,p)}{\sqrt{A_s^2(p_0,p)+A_m^2(p_0,p)}}\left(\frac{1}{D_+(p_0,p)} - \frac{1}{D_-(p_0,p)} \right)\right]\right\}.
	\ee
%%%%%%%%%%%

\noindent The total contribution can be written as a sum of the QP and LD contribution as
\be
\chi_{ff}(T,m_f)=	\chi_{ff}^{QP}(T,m_f)+	\chi_{ff}^{LD}(T,m_f).
\ee
Now, second-order BNS can be related with second-order QNSs as
\be
\chi_2^B(T,\bm{m})=\frac{1}{9}\Big[\chi_{uu}+\chi_{dd}+\chi_{ss}+2\chi_{ud}+2\chi_{ds}+2\chi_{us}\Big].
\label{gen_chi2}
\ee
At a vanishing chemical potential, second-order off-diagonal susceptibilities are zero. Additionally, we consider the masses of $u$ and $d$ quarks as the same. So, Eq.~(\ref{gen_chi2})
becomes
\be
\chi_2^B(T,\bm{m})=\frac{1}{9}\Big[2\chi_{uu}+\chi_{ss}\Big].
\ee
\begin{figure}[tbh!]
	\subfigure{
\includegraphics[width=8.5cm]{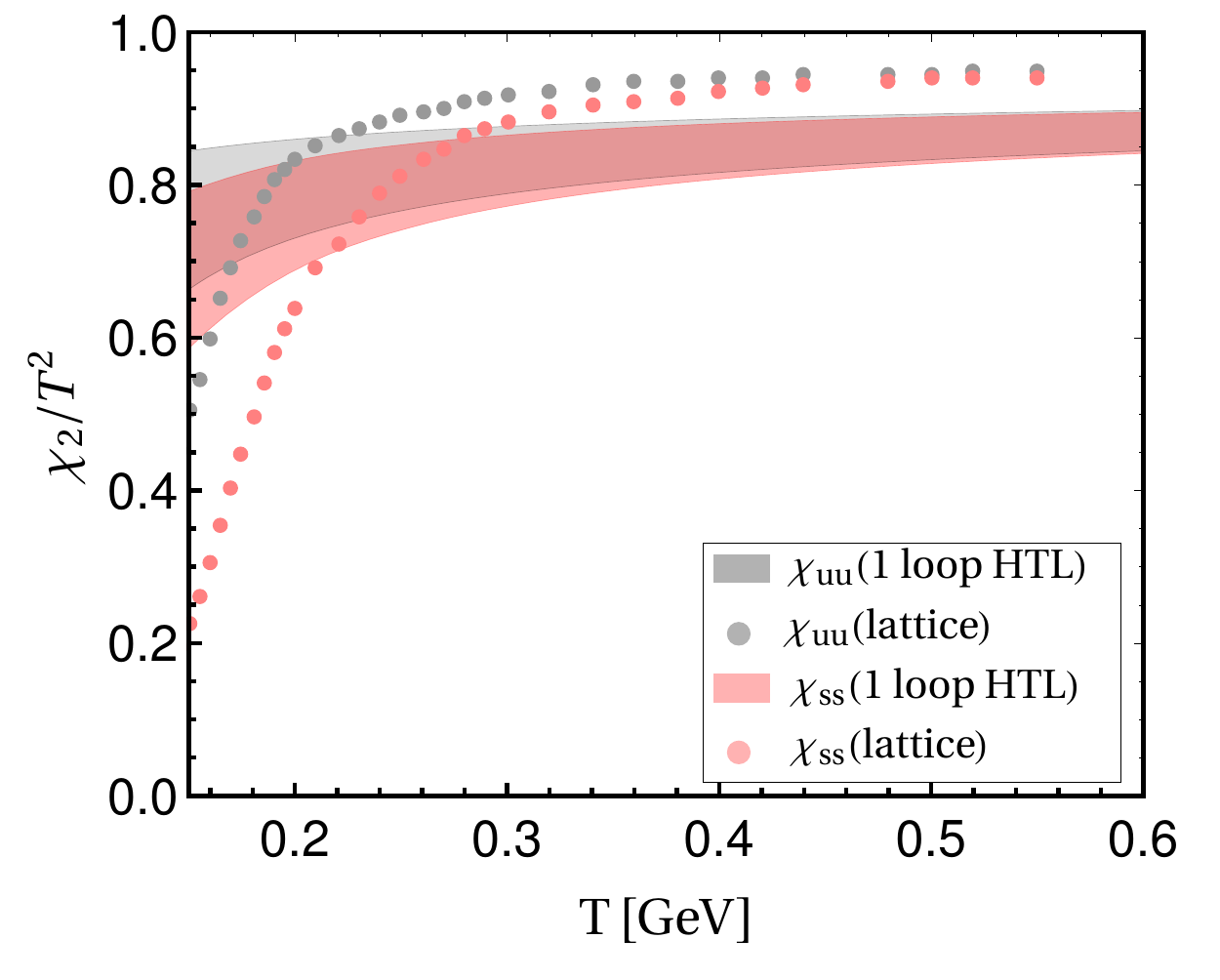}}
\subfigure{\includegraphics[width=8.5cm]{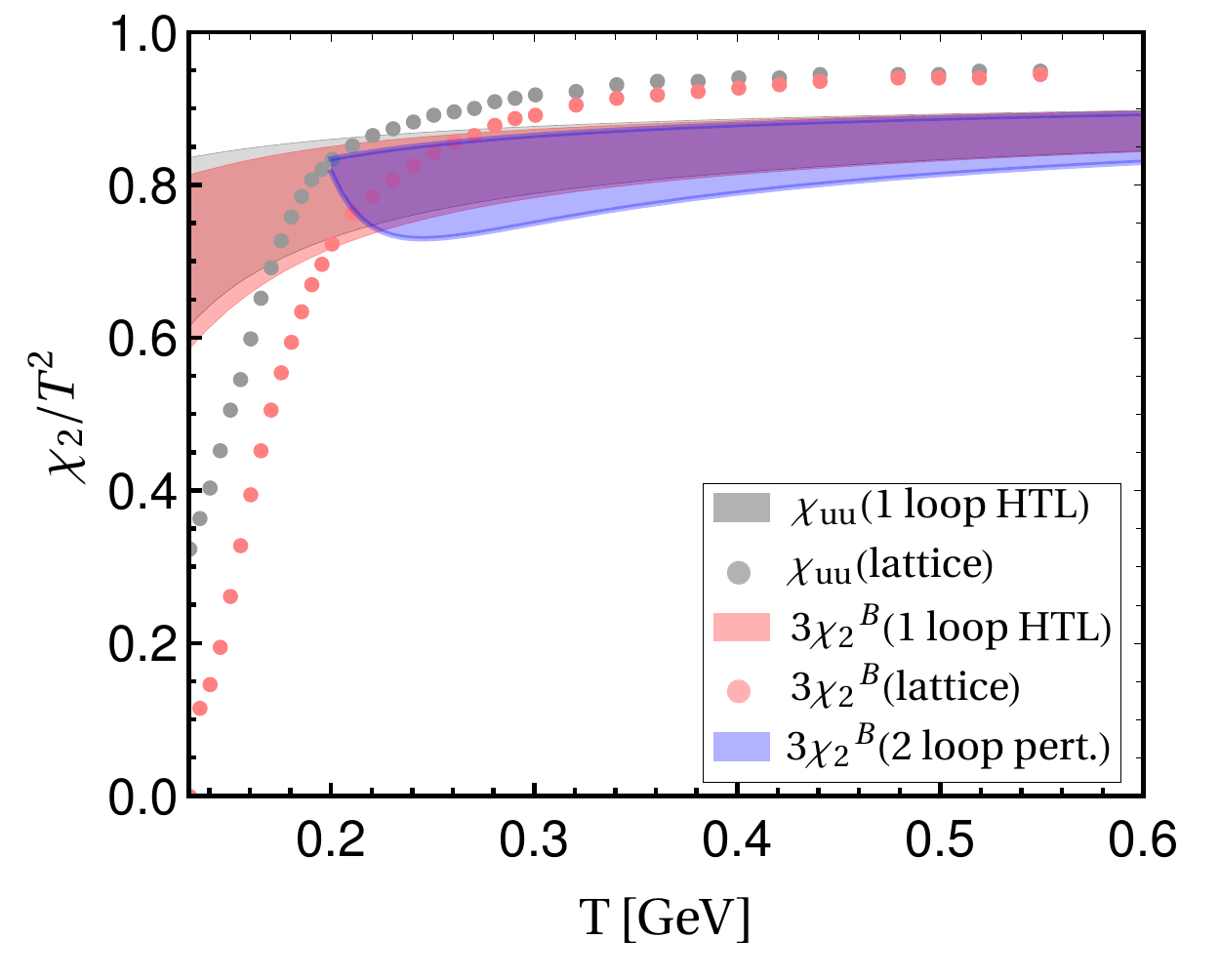}}
\caption{Left: Comparison of the one-loop HTLpt results for the scaled light quark and strange quark number susceptibilities with the lattice results~\cite{Bellwied:2015lba}. Right: Comparison of the one-loop HTLpt results for the scaled light quark and baryon number susceptibilities with the lattice~\cite{Bellwied:2015lba} and two-loop perturbative results~\cite{Graf:2015tda}.}
\label{fig:qns}
\end{figure} 
\end{widetext}
% % % % % % % % % % % % % % % % 
\noindent In Fig.~\ref{fig:qns}, we compare the second-order light and strange quark susceptibilities (left panel) and also light quark susceptibility and three times the baryon number susceptibility (right panel) with available lattice data from the Wuppertal- Budapest (WB) group~\cite{Bellwied:2015lba} and also with two-loop perturbative results~\cite{Graf:2015tda}. Lattice QCD data have been obtained using the four-level-smeared (4stout) staggered action. In that calculation, temperature-dependent quark masses, obtained by simulating the finite temperature ensembles with physical quark masses, have been used. The two-loop perturbative second-order BNS~\cite{Graf:2015tda} has been calculated considering chemical potential-dependent renormalization scale $\Lambda$ for both quark and gluon diagrams, which might not be the appropriate way. To introduce a chemical potential dependence to the renormalization scale, one needs to define two renormalization scales, namely, $\Lambda_q$ (for quarks) and $\Lambda_g$ (for gluons). $\Lambda_q$ will be chemical potential dependent, whereas $\Lambda_g$ will be chemical potential independent. By doing so, one will get a different BNS plot than what is shown in Fig.~\ref{fig:qns}.

The band in Fig.~\ref{fig:qns} indicates the sensitivity of susceptibilities with the renormalization scale, and we use the values of the renormalization scale $\pi T<\Lambda<4\pi T$. In this plot, we use the temperature-dependent strange quark mass as given in Eq.~(\ref{ms}). As the effect of the light quark mass is almost invisible in the QNS or BNS plot, we take those as zero. The resulting quark and baryon number susceptibilities are $(5\textendash10)\%$ lower than the WB group lattice data at high temperature. In the low-temperature region, we have gotten the same trend as the LQCD result, that strange quark number susceptibility and three times the baryon-number susceptibility are lower than light quark susceptibility. 
Additionally, the one-loop HTLpt BNS matches very well with two-loop perturbative results for temperature $T\geq 200$ MeV, though the renormalization scale-dependent band is smaller for the one-loop HTL BNS.
% % % % % % % % % % % % % 
\section{Conclusion and outlook}
\label{sec:conclusion}
\noindent In this article, we have calculated all the QCD $N$-point functions such as the quark and gluon propagators and three- and four-point quark-gluon vertices at a finite quark mass within the hard thermal loop approximation. The resulting $N$-point functions satisfy the WI with corresponding $(N-1)$-point functions. We have also calculated second-order quark and baryon number susceptibility within the hard thermal loop approximation. In massless perturbative calculations, one cannot distinguish between second-order quark and baryon number susceptibilities. Considering a finite strange quark mass, for the first time in any resummed perturbative framework, we are able to distinguish them.

% % % % % % % % % % % % 
\noindent Looking to the future, we want to extend the current calculation to the next-to-next-to-leading order (NNLO) to study all the thermodynamical quantities. As the current result are $(5\textendash10)\%$ lower than the LQCD data at high temperature, we expect good agreement with LQCD data in NNLO within the HTLpt in the presence of finite quark masses. In addition, we will also be able to differentiate between two off-diagonal fourth-order susceptibilities $(\chi_{uudd}$ and $\chi_{uuss})$ in NNLO at a nonvanishing strange quark mass.

% We expect that inclusion of finite quark masses to the NNLO thermodynamics will reduce the renormalization scale dependent band and will also improve the low temperature behavior of the existing massless results~\cite{Haque:2014rua}.
% % % % % % % % % % 
\section{Acknowledgements}
\label{sec:ack}
\noindent The author was supported by a postdoctoral research fellowship from the Alexander von Humboldt
Foundation, Germany. The author acknowledges A.~Bandyopadhyay, S.~Chakraborty, and M.~G.~Mustafa for useful discussion and careful reading of the article.

\appendix

\section{Gluon propagator}
\label{sec:gluon_prop}
\begin{figure}[tbh ]
	\includegraphics[width=5cm]{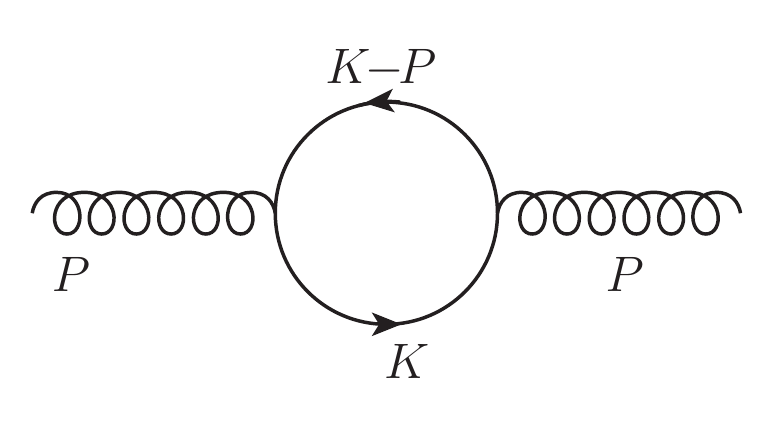}
	\caption{Gluon self-energy diagram}
	\label{fig:gluon_self}
\end{figure}

\noindent
The one-loop Feynman diagram for the gluon self-energy consisting the quark loop is illustrated in Fig.~\ref{fig:gluon_self}.

\noindent The contribution from the Feynman diagram in Fig~\ref{fig:gluon_self} to the  gluon self-energy in Minkowski space can be written as
\be
&&\hspace{-.5cm}\Pi_{\mu\nu}^{ab,f}(P, \bm{m})\nn
&=& (-1)(ig)^2(t^a)_{ij}(t^b)_{ji} \sum_{f}\int\!\frac{d^4K}{(2\pi)^4}\nn
&\times&\textsf{Tr}\left [\gamma_\mu S_0(K)\gamma_\nu S_0(K-P)\right ]\nn
&=& g^2\textsf{Tr}[t^at^b] \sum_{f}\int\frac{d^4K}{(2\pi)^4}\textsf{Tr}\left [\gamma_\mu S_0(K)\gamma_\nu S_0(Q)\right ]\nn
&=& i\ \Pi^f_{\mu\nu}(P,\bm{m}) \delta^{ab},\ee
with $S_0(K)=i/(\slashed{K}-m_f)$ and $Q=K-P$.
%%%%%%%%%
So, the gluon self-energy tensor $\Pi^f_{\mu\nu}(P,\bm{m})$ is
%%%%%%%%%%%%%%%%%%%%%%%%%%%%%
\be
&&\hspace{-.8cm}\Pi^f_{\mu\nu}(P,\bm{m})\nn
&=&-\frac{ig^2}{2}\sum_{f}\!\int\!\!\frac{d^4K}{(2\pi)^4}\textsf{Tr}\left [\gamma_\mu S_0(K)\gamma_\nu S_0(Q)\right]\nn
&\simeq&\frac{ig^2}{2}\sum_{f}\int\frac{d^4K}{(2\pi)^4}\left [8K_\mu K_\nu - 4 g_{\mu\nu}\(K^2-m_f^2\)\right]\nn
&\times&\tilde{\Delta}(K)\tilde{\Delta}(Q)
\ee
% % % % % % % % % % % % 

Now, the time-time component of the self-energy tensor becomes
\be
\hspace*{0cm}&&\hspace*{-.5cm}\Pi_{00}^f(P,\bm{m})\nn
&=&\frac{4ig^2}{2}\sum_{f}\!\!\int\!\!\frac{d^4K}{(2\pi)^4}\(2k_0^2-K^2+m_f^2\)\tilde{\Delta}(K)\tilde{\Delta}(Q)\nn
&=&\frac{4ig^2}{2}\!\sum_{f}\!\!\int\!\!\frac{d^4K}{(2\pi)^4}\Big[\tilde{\Delta}(Q)+2\(k^2+m_f^2\)\tilde{\Delta}(K)\tilde{\Delta}(Q)\Big]\nn
&\simeq& 2ig^2\!\sum_{f}\!\!\int\!\!\frac{d^4K}{(2\pi)^4}\Big[\tilde{\Delta}(K)+2\(k^2+m_f^2\)\tilde{\Delta}(K)\tilde{\Delta}(Q)\Big].\nn
\label{pi00_def}
\ee

\noindent Similarly, the trace of the gluon self-energy tensor is
\be
&&\hspace{-0.5cm}{\Pi_\mu^\mu}^f(P,\bm{m})\nn
&\simeq&-4ig^2\sum_{f}\!\int\!\frac{d^4K}{(2\pi)^4}\Big[\tilde{\Delta}(K)-m_f^2\tilde{\Delta}(K)\tilde{\Delta}(Q)\Big].
\label{pimumu_def}
\ee
The frequency sum for the first term in Eqs.~(\ref{pi00_def}) and ~(\ref{pimumu_def}) can be evaluated easily as
\be
&&\hspace*{-1cm}i\int\frac{d^4K}{(2\pi)^4}\tilde{\Delta}(K)\nn
&=&-T\sum_{n}\int\frac{d^3k}{(2\pi)^3}\frac{1}{k_0^2-k^2-m_f^2}\nn
&=&-\int\frac{k^2dk}{2\pi^2}\frac{n_F^+(E_k) + n_F^-(E_k)}{2E_k},
% &=&i\int\frac{k^2dk}{4\pi^2}\frac{n_F}{E_k},
\label{sum1}
\ee
where
\be
E_k&=&\sqrt{k^2+m_f^2},\\
n_F^\pm(E_k)&=&\frac{1}{\exp[\beta\(E_k\mp\mu_f\)]+1}.
\ee
%%%%%%%%%%%%%%%
%\be
%\int\frac{d^4K}{(2\pi)^4}\tilde{\Delta}(K)\tilde{\Delta}(Q)&=&iT\sum_{n}\int\frac{d^3k}{(2\pi)^3}\frac{1}{k_0^2-k^2-m_f^2}\frac{1}{(k_0-p_0)^2-|{\bf k-p}|^2-m_f^2}\nn
%&=&,
%\ee
We calculate the second Matsubara sum of Eqs.~(\ref{pi00_def}) and~(\ref{pimumu_def}) using the mixed representation prescribed by Pisarski~\cite{Pisarski:1987wc}.

\begin{widetext}
 Using the relations in Eqs.~(\ref{Delta_tilde}) and~(\ref{Delta_F}), the following Matsubara sum can be performed as
\be
&&\hspace{-.5cm}T\sum\limits_{k_0}\tilde{\Delta}(K)\tilde{\Delta}(P-K) = \int\limits_0^\beta d\tau 
e^{p_0\tau} \Delta_F(\tau,E_k)\Delta_F(\tau,E_{pk})\nn
&=& -\sum\limits_{s_1,s_2=\pm 1} 
\frac{s_1s_2\left[1-n_F^-(s_1E_k) - n_F^+(s_2E_{pk})\right]}{4E_kE_{pk}
	\left(p_0-s_1E_k-s_2E_{pk}\right)}\nn
&=&\frac{1}{4 E_k E_{pk}}\Bigg[\left(\frac{n_F^+(E_k)-n_F^+(E_{pk})}{p_0+E_k-E_{pk}}-\frac{n_F^-(E_k)-n_F^-(E_{pk})}{p_0-E_k+E_{pk}}\right)\nn
&&
\hspace{1.2cm}-\ \left(\frac{1-n_F^-(E_k)-n_F^+({E_pk})}{p_0-E_k-E_{pk}} - \frac{1-n_F^+(E_k)-n_F^-(E_{pk})}{p_0+E_k+E_{pk}}\right)\Bigg].
\label{sum2}
\ee
\end{widetext}
where
\be
E_k=\sqrt{k^2+m_f^2}\ ,\\
E_{pk}=\sqrt{|\mathbf{p-k}|^2+m_f^2}\ .
\ee
In the HTL approximation, 
\be
E_{pk}=\sqrt{|\bm{p-k}|^2+m_f^2}\approx E_k -v\,\bm{\hat{k}\cdot p},\nn
n_F^\pm(E_{pk})\approx n_F^\pm(E_k) - \bm{\hat{k}\cdot p} \frac{dn_F^\pm(E_k)}{dk},
\label{htl_approx}
\ee
with $v=k/E_k$.

\noindent Using the HTL approximations [Eq.~(\ref{htl_approx})], the Matsubara sum in Eq.~(\ref{sum2}) can be written as
\be
&&\hspace{-.8cm}T\sum\limits_{k_0}\tilde{\Delta}(K)\tilde{\Delta}(P-K)= \frac{1}{4 E_k^2}\Bigg[-\frac{n_F^+(E_k)+n_F^-(E_k)}{E_k}\nn
&&\hspace{-.4cm}+\frac{dn_F^+(E_k)}{dk}\frac{\bm{\hat{k}\cdot p}}{p_0+v\,\bm{\hat{k}\cdot p}}
-\frac{dn_F^-(E_k)}{dk}\frac{\bm{\hat{k}\cdot p}}{p_0-v\,\bm{\hat{k}\cdot p}}\Bigg].\ 
\ee

\noindent So, Eq.~(\ref{pi00_def}) becomes
\be
&&\hspace{-.5cm}\Pi_{00}^f(p_0,p,\bm{m})\nn
&=&2ig^2\sum_{f}\int\frac{d^4K}{(2\pi)^4}\ \Big[\tilde{\Delta}(K)+2E_k^2\tilde{\Delta}(K)\tilde{\Delta}(Q)\Big]\nn
&=&-2g^2\sum_{f}\!\!\int\!\frac{k^2dk}{4\pi^2}\frac{d\Omega}{4\pi}\nn
&&\hspace{0cm}\times\frac{d}{dk}\Big[n_F^+(E_k)+n_F^-(E_k)\Big]\frac{\bm{\hat{k}\cdot p}}{p_0+v\,\bm{\hat{k}\cdot p}}\nn
&=&-\frac{g^2}{2\pi^2}\!\int\! k^2dk\frac{E_k}{k}\frac{d}{dk}\Big[n_F^+(E_k)+n_F^-(E_k)\Big]\nn
&&\hspace{0.6cm}\times\left[1-\frac{p_0}{2p v}\log\frac{p_0+pv}{p_0-pv}\right].
\label{pi00_k_deriv}
\ee
Equation~(\ref{pi00_k_deriv}) can also be simplified as
\be
&&\hspace{-.5cm}\Pi_{00}^f(p_0,p,\bm{m})\nn
&=&\frac{g^2}{2\pi^2}\sum_{f}\!\!\int \frac{k^2dk}{E_k}\Big[n_F^+(E_k)+n_F^-(E_k)\Big]\nn
&&\times\left[1+\frac{p_0^2-p^2}{p_0^2-p^2v^2}-\frac{p_0}{p v}\log\frac{p_0+pv}{p_0-pv}\right].
\ee

\noindent Similarly, the trace of the gluon self-energy tensor in Eq.~(\ref{pimumu_def}) can be written using the Matsubara sums in Eqs.~(\ref{sum1}) and~(\ref{sum2}) as
\be
&&\hspace{-0.5cm}{\Pi_\mu^\mu}^f(p_0,p,\bm{m})\nn
&=&-4ig^2\sum_{f}\int\frac{d^4K}{(2\pi)^4}\ \Big[\tilde{\Delta}(K)-m_f^2\tilde{\Delta}(K)\tilde{\Delta}(Q)\Big]\nn
&=&\frac{g^2}{\pi^2}\sum_{f}\int k^2dk\frac{d\Omega}{4\pi}\Bigg[\frac{n_F^+(E_k)+n_F^-(E_k)}{E_k}\nn
&&+\ \frac{m_f^2}{2E_k^2}\Bigg\{\frac{n_F^+(E_k)+n_F^-(E_k)}{E_k}\nn
&& \ -\ \frac{E_k}{k}\frac{d}{dk}\Big[n_F^+(E_k)+n_F^-(E_k)\Big]\left(1-\frac{p_0}{p_0+p v x}\right)\Bigg\}\nn
%%%%%%%%%%%%%%%%%%%%%%%
&=&\frac{g^2}{\pi^2}\sum_{f}\int k^2dk\Bigg[\frac{n_F^+(E_k)+n_F^-(E_k)}{E_k}\left(1+\frac{m_f^2}{2E_k^2}\right)\nn
&&-\ \frac{m_f^2}{2kE_k}\frac{d}{dk}\Big[n_F^+(E_k)+n_F^-(E_k)\Big]\nn
&&\ \times\left(1-\frac{p_0}{2p v}\log\frac{p_0+pv}{p_0-pv}\!\right)\!\!\Bigg]\!.
%%%%%%%%%%%%%%%%%%%%%%%
\label{pimumu_k_deriv}
\ee
Equation~(\ref{pimumu_k_deriv}) can also be simplified as
\be
{\Pi_\mu^\mu}^f(p_0,p,\bm{m})&=&\frac{g^2}{\pi^2}\sum_{f}\!\int\!\! \frac{k^2dk}{E_k}\(n_F^+(E_k)+n_F^-(E_k)\)\!\nn
&\times&\Bigg[1
+\frac{m_f^2}{2E_k^2}\frac{p_0^2-p^2}{p_0^2-p^2v^2}\Bigg].\qquad
\ee
Pure Yang-Mills contributions to the gluon self-energy do not get affected by the inclusion of finite quark masses. So, the time-time and $\mu\mu$ components of the total gluon self-energy tensor from the fermion as well as pure Yang-Mills diagrams is
\be
&&\hspace{-0.5cm}\Pi_{00}(p_0,p,\bm{m})\nn
&=&\frac{g^2T^2C_A}{3}\left[1-\frac{p_0}{2p}\log\frac{p_0+p}{p_0-p}\right]\nn
&&+\frac{g^2}{2\pi^2}\sum_{f}\int \frac{k^2dk}{E_k}\Big[n_F^+(E_k)+n_F^-(E_k)\Big]\nn
&&\times\left[1+\frac{p_0^2-p^2}{p_0^2-p^2v^2}-\frac{p_0}{p v}\log\frac{p_0+pv}{p_0-pv}\right]
\label{pi00_total}
\ee
and
\be
&&\hspace{-0.5cm}\Pi_\mu^\mu(p_0,p,\bm{m})\nn
&=&\frac{g^2T^2C_A}{3}+\frac{g^2}{\pi^2}\sum_{f}\!\int\! \frac{k^2dk}{E_k}\Big[n_F^+(E_k)+n_F^-(E_k)\Big]\nn
&\times&\left[1+\frac{m_f^2}{2E_k^2}\ \frac{p_0^2-p^2}{p_0^2-p^2v^2}\right].
\label{pimumu_total}
\ee
Note that, unlike the massless case, gluonic and quark contributions cannot combined together.

\noindent Now the one-loop gluon self-energy can be decomposed in terms of two independent and mutually transverse second rank projection tensors as
\be
\Pi_{\mu\nu}=\Pi_T A_{\mu\nu} + \Pi_L B_{\mu\nu},
\ee
with
\be
A_{\mu\nu}&=& g_{\mu\nu} -\frac{P_\mu P_\nu}{P^2}-B_{\mu\nu},\\
B_{\mu\nu}&=&-\frac{P^2}{p^2}\left(u_\mu-\frac{p_0P_\mu}{P^2}\right)\left(u_\nu-\frac{p_0P_\nu}{P^2}\right).
\ee
So, the longitudinal and transverse parts of gluon self-energy tensor can be obtained from Eqs.~(\ref{pi00_total}) and~(\ref{pimumu_total}) as
\be
&&\hspace{-1cm}\Pi_L(p_0,p,\bm{m})\nn
&=&-\frac{p_0^2-p^2}{p^2}\Pi_{00}(p_0,p,\bm{m})\nn
&=&-\frac{p_0^2-p^2}{p^2}\Bigg[\frac{g^2T^2C_A}{3}\left(1-\frac{p_0}{2p}\log\frac{p_0+p}{p_0-p}\right)\nn
&+&\frac{g^2}{2\pi^2}\sum_{f}\int \frac{k^2dk}{E_k}\Big(n_F^+(E_k)+n_F^-(E_k)\Big)\nn
&\times&\left(1+\frac{p_0^2-p^2}{p_0^2-p^2v^2}-\frac{p_0}{p v}\log\frac{p_0+pv}{p_0-pv}\right)\Bigg]
\label{piL_total}
\ee
and
\be
&&\hspace{-0.5cm}\Pi_T(p_0,p,\bm{m})\nn
&=&\frac{1}{2}\left[\Pi_\mu^\mu(p_0,p,m)-\Pi_L(p_0,p,m)\right]\nn
&=&\frac{g^2T^2C_A}{3}\left[1+\frac{p_0^2-p^2}{p^2}\left(1-\frac{p_0}{2p}\log\frac{p_0+p}{p_0-p}\right)\right]\nn
&&+\frac{g^2}{2\pi^2}\sum_{f}\!\int\! \frac{k^2dk}{E_k}\Big[n_F^+(E_k)+n_F^-(E_k)\Big]\nn
&&\hspace{0.2cm}\times\left(\frac{2p_0^2}{p^2}-\frac{p_0^2-p^2}{p^2}\frac{p_0}{p v}\log\frac{p_0+pv}{p_0-pv}\right).
\label{piT_total}
\ee
%%%%%%%%%%%%%%%%%
\subsection{Debye mass}
\noindent The Debye mass in QCD is obtained as
\be
m_D^2=\Pi_L(p_0=0,p\rightarrow 0,\bm{m}).
\ee
At vanishing quark masses, the Debye mass is
\be
m_D^2(T,\mu)=\frac{g^2T^2}{3}\left[C_A+\frac{N_f}{2}\left(1+\frac{3\hat{\mu}^2}{\pi^2T^2}\right)\right],
\ee
where $\mu$ a is common chemical potential for all the quark flavors.
Now, the Debye mass with finite quark masses can be obtained from Eq.~(\ref{piL_total}) as
\be
&&\hspace{-0.8cm}m_D^2(T,\bm{\mu}, \bm{m})
=\Pi_L(p_0=0,p\rightarrow 0,\bm{m})\nn
&=&\frac{g^2T^2C_A}{3} +\frac{g^2}{2\pi^2}\sum_{f=1}^{N_f}\nn
&\times&\int\limits_0^\infty \frac{k\,dk}{v}\(1+v^2\)\Big[n_F^+(E_k)+n_F^-(E_k)\Big].
\ee
%%%%%%%%%%%%
\noindent  The plasma frequency is the oscillation frequency for vanishing wave
vectors, namely, spatially uniform oscillations~\cite{Bellac:2011kqa} and can be expressed as
\be
\omega_p^2&=&\frac{g^2T^2C_A}{9} + \frac{g^2}{2\pi^2}\sum_{f=1}^{N_f}\nn
&\times&\int\limits_0^\infty dk k v\left(1-\frac{v^2}{3}\right)\Big[n_F^+(E_k)+n_F^-(E_k)\Big].
\ee
Unlike massless case, the Debye mass and plasma frequency are not proportional to each other.
\begin{figure}[tbh]
\includegraphics[width=8cm]{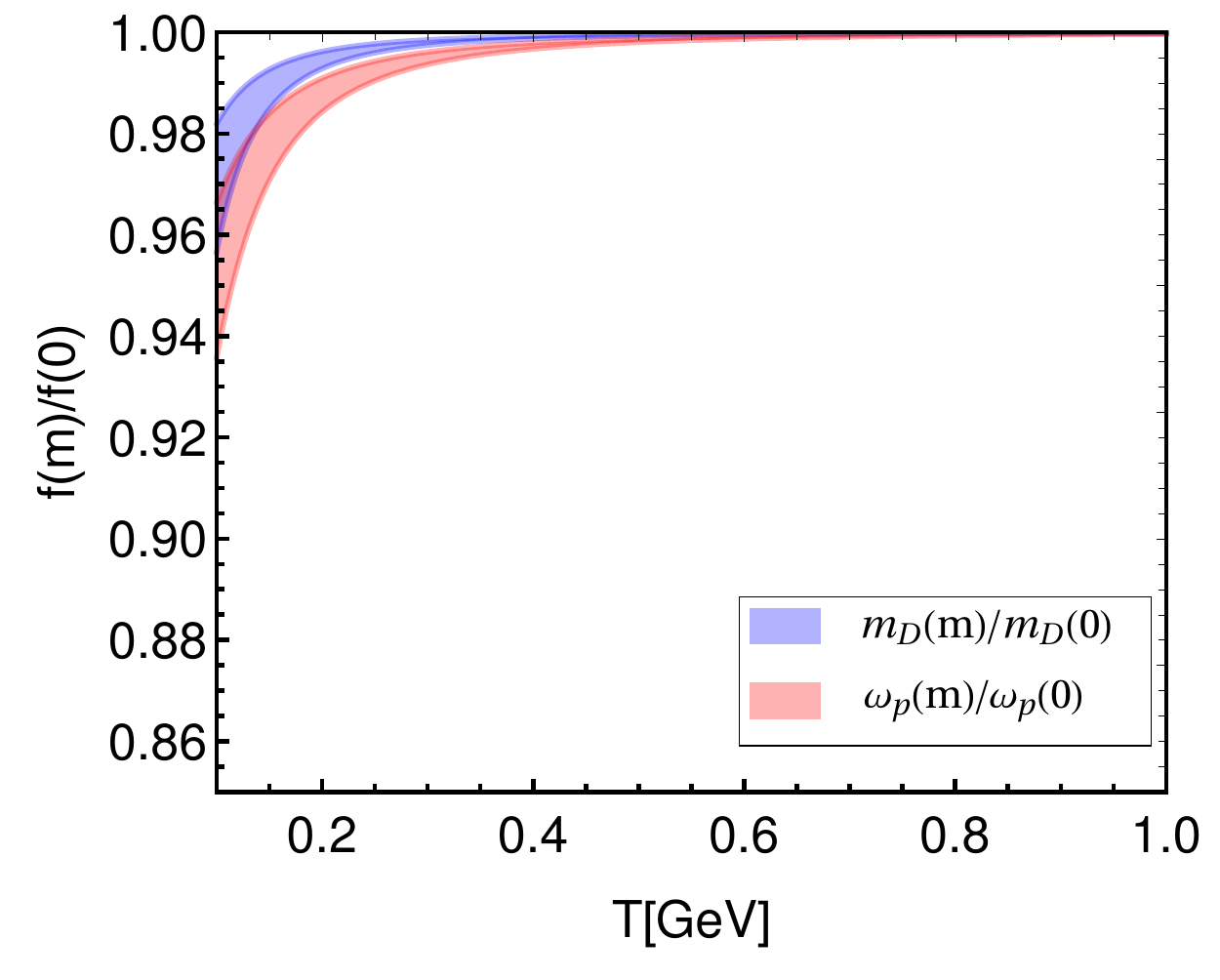}
\caption{Debye mass and plasma frequency plotted with the temperature}
\label{fig:mD_wp}
\end{figure}
%%%

In Fig.~\ref{fig:mD_wp}, we plot quark mass-dependent Debye mass and plasma frequency, scaled with corresponding massless values, with the temperature. We use the strange quark mass from Eq.~(\ref{ms}), whereas light-quark mass as zero. The renormalization scale $(\Lambda)$ dependence band appears due to the choice of a scale -dependent strange quark mass.
%%%%%%%%%%%%%%%%%%%%%%%
\subsection{Dispersion relation}
\label{subsec:disp}
\begin{figure}[tbh]
% 	\subfigure{
% 		\includegraphics[width=8cm]{disp_real_mass.pdf}}
	\subfigure{
		\includegraphics[width=6.7cm]{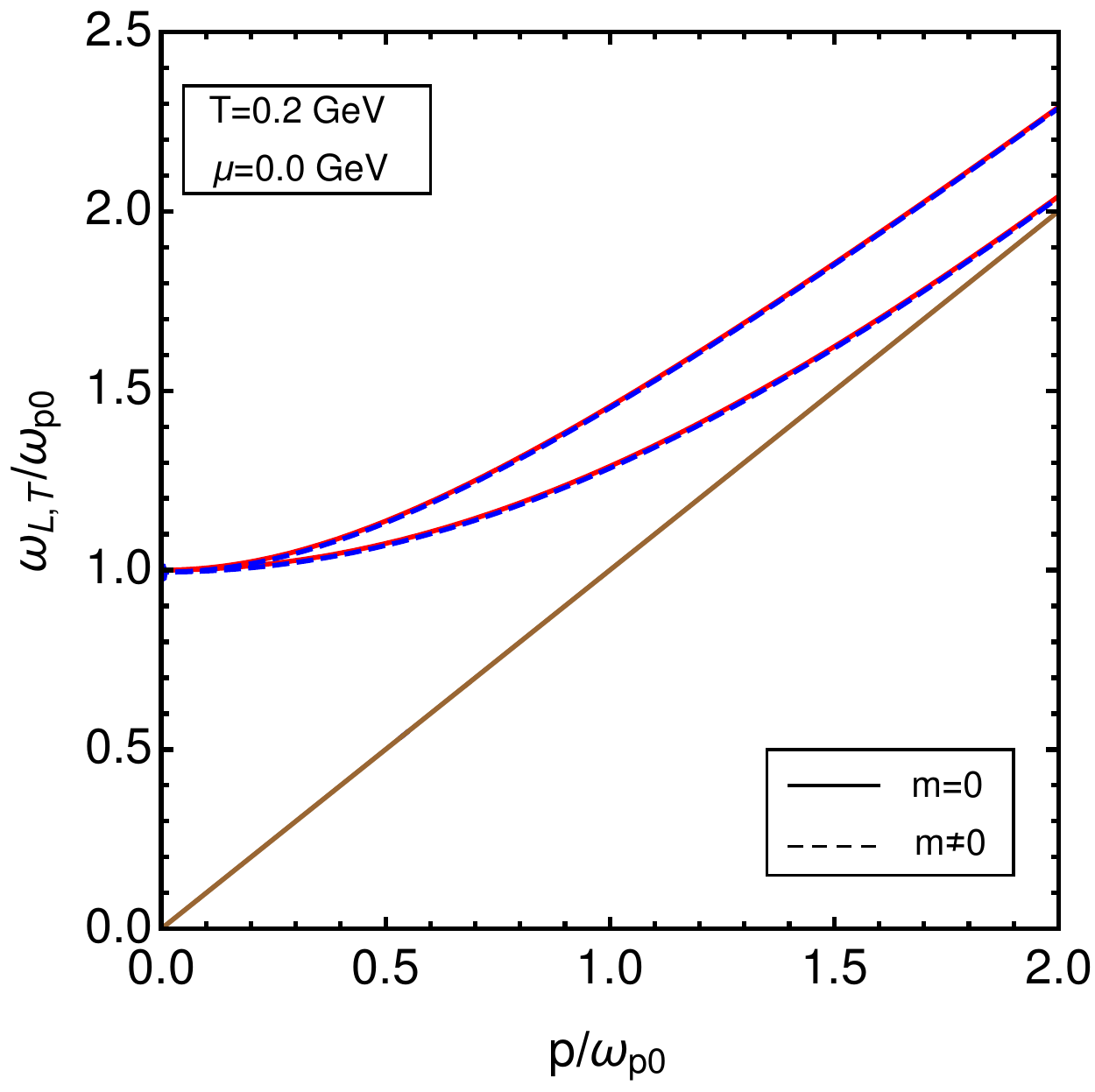}}
	\caption{Gluon dispersion plots scaled with the massless plasma frequency of the plasma}
\label{fig:disp_gluon}
\end{figure}

\noindent The in-medium gluon propagator can be written as
\be
\mathcal{D}_{\mu\nu} &=& \frac{\xi P_\mu P_\nu}{P^4} +\frac{ A_{\mu\nu}}{P^2-\Pi_T} + \frac{B_{\mu\nu}}{P^2-\Pi_L}\nn
&=& \frac{\xi P_\mu P_\nu}{P^4} +\frac{ A_{\mu\nu}}{\mathcal{D}_T} + \frac{B_{\mu\nu}}{\mathcal{D}_L}.
\ee
%%%%%%%%%%
\noindent The zeros of the denominators give the dispersion laws as $\omega_L$ and $\omega_T$ as plotted in Fig.~\ref{fig:disp_gluon}.
\noindent In Fig.~\ref{fig:disp_gluon} we plot longitudinal and transverse dispersion relations scaled with the massless plasma frequency.

\noindent At small momentum $\left(p\ll T\right)$, it is possible to find approximate analytic solution of $\omega_{L,T}$ as
\be
\omega_L^2&=&\omega_p^2+\frac{p^2}{\omega_p^2}a_{L}+\frac{p^4}{\omega_p^4}b_{L}+\mathcal{O}\left(p^6\right),\\
%%%%%%%%%%%%%%%%%%
\omega_T^2&=&\omega_p^2+\frac{p^2}{\omega_p^2}a_{T}+\frac{p^4}{\omega_p^4}b_{T}+\mathcal{O}\left(p^6\right),
\ee
where
\be
a_{L}&=&\frac{g^2T^2C_A}{15}\
+\frac{g^2}{2\pi^2}\sum_{f=1}^{N_f}\nn
&\times&\!\int\limits_0^\infty\! dk k v^3\left(1-\frac{3v^2}{5}\right)\Big[n_F^+(E_k)+n_F^-(E_k)\Big]\!,\,\\
b_{L}&=&\frac{g^2T^2C_A}{21}-\frac{a_L^2}{\omega_p^2}+ \frac{g^2}{2\pi^2}\sum_{f=1}^{N_f}\nn
&\times&\!\int\limits_0^\infty\! dk k v^5\left(1-\frac{5v^2}{7}\right)\Big[n_F^+(E_k)+n_F^-(E_k)\Big]\!,\,\\
a_{T}&=&\omega_p^2+\frac{a_{L}}{3},\\
b_T&=&\frac{b_L}{5}-\frac{a_L}{3}+\frac{4}{45}\frac{a_L^2}{\omega_p^2}.
\ee

\noindent From Figs.~\ref{fig:mD_wp} and~\ref{fig:disp_gluon}, we conclude that the finite mass effect to the gluon collective excitations is negligible, and one may consider the effective gluon propagator is the same as in the massless case to calculate various physical observables.
% % % % % % % % % % % % % % % % % % % % % % % % 
% % % % % % % % % % % % % % % % % % % % % % % % % %
% % % % % % % % %   References   % % % % % % % % % % 
% % % % % % % % % % % % % % % % % % % % % % % % % %

\end{document}